\documentclass[preprint,12pt]{elsart}

\usepackage{amsmath}
\usepackage{bm} 
\usepackage{graphicx} 
\usepackage{epstopdf}

\journal{Surface Science}
\begin{document} 
\begin{frontmatter}
\title{Diffusion of Cu adatoms and dimers on Cu(111) and Ag(111) surfaces.} 

\author{Marcin Mi{\'n}kowski and Magdalena A Za{\l}uska--Kotur}

\address{Institute of Physics, Polish
Academy of Sciences, Al.~Lotnik{\'o}w 32/46, 02--668 Warsaw, Poland}
\ead{minkowski@ifpan.edu.pl, zalum@ifpan.edu.pl} 


\begin{abstract}
Diffusion of Cu adatoms and dimers on Cu(111) and Ag(111) surfaces is analyzed basing on ab-initio surface potentials. Single adatom diffusion is compared with dimer diffusion on both surfaces. Surface geometry makes the adatoms jump alternately between two states in the same way in both systems, whereas dimers undergo more complex diffusion process that combines translational and rotational motion. Small difference in the surface lattice constant between Cu and Ag crystals results in a completely different energy landscape for dimer jumps. As an effect the character of diffusion process changes. Homogeneous Cu dimer diffusion is more difficult and dimers rather rotate within single surface cell, whereas diffusion over Ag surface is faster and happens more smoothly. Temperature dependence of diffusion coefficient and its parameters: energy barrier and prefactor is calculated and compared for both surfaces.
 \end{abstract}
\end{frontmatter}
\section{Introduction}
Surface diffusion coefficient is one of the factors that decide about  the crystal growth process. In order to understand formation of nanostructures it is necessary to understand the underlying microscopic processes, such as adsorption, desorption and diffusion. A lot of studies focus on dynamics of a single adatom \cite{gomer,liu,tringides}. However, 
diffusion of small clusters, e.g., dimers and trimers, which
occurs with lower barriers \cite{ehrlich,Michely1,Evans1,kellog}, can be also significant  and can affect island formation.In particular on  metallic surfaces even diffusion of larger clusters   can be very effective\cite{Evans1,kellog}.

Diffusive motion of atoms and clusters at the surface is studied experimentally by scanning tunneling microscopy (STM) and field ion microscopy (FIM), both used for direct imaging of individual adatoms \cite{Morgenstern1,Morgenstern2,Repp1,bogicevic}. With the help of other techniques, such as field emission microscopy (FEM), photoemission electron microscopy  diffusion coefficient can be measured.

As for theoretical approaches, the first step is always calculation of the adiabatic potential, that means all the adsorption sites and energy barriers between them \cite{Marinica1,karim,wu,Wang1,Hayat1}. Then these data are used to analyze the diffusive motion. It can be done by means of simulation methods, such as kinetic Monte Carlo (kMC) or molecular dynamics \cite{Wang1,Hayat1,Michailov1,Michailov2,Michailov3,Flores1}. The latter solves the equations of motion for adatoms in a given potential. Their results are, however, limited to shorter time scales than those obtained in kMC simulations. Both these methods must be applied for specified potential or jump rates and at specified temperature. By changing any of these, one has to perform the simulation again. Below we derive analytic expressions for diffusion coefficients for adatoms and dimers based on the existing picture of the potential energy landscape.
Otherwise than MC or molecular dynamic calculations, analytic formulas express diffusion coefficient as a function of temperature and other model parameters \cite{ala,penev,penev2,haus}. It gives much more insight into the real physics of the system. Moreover, if new, more accurate data about the energy landscape are obtained, it is possible to improve the diffusion coefficient values on using the same formulas. Among theoretical approaches there are methods based on Langevin equation, Fokker-Planck equation or path integral formalism \cite{ala}. Our approach, on the other hand, is built on the master equation. Similarly to the method developed by Titulaer and Deutch \cite{Titulaer1}, which was later applied in \cite{Sanchez1,Salo1} for metal surfaces, we find the diffusive eigenvalue of the rate matrix. However, we do that by means of a variational method.

In this work we calculate the diffusion coefficient for Cu monomer and Cu dimer on Cu(111) and Ag(111) surfaces. In such a way dynamics of homo- and heteroepitaxial system is compared for the same adatom. Such systems have been investigated both experimentally \cite{Morgenstern1,Morgenstern2,Repp1} and theoretically \cite{Marinica1,karim,wu,Wang1,Hayat1,Flores1}. In addition to diffusion of a dimer, there have been also studies concerning adatom clusters of larger size \cite{karim,wu,Wang1,Flores1,signor}.  Hetero-diffusion appears to be easier with lower activation barrier and homo-diffusion is dominated by relatively high energy barriers between neighboring cells, whereas rotation inside each cell can run freely.

We  calculate diffusion coefficient using adsorption energies and energetic barriers for jumps between sites obtained in Refs.\cite{karim,Hayat1} by means of ab-initio calculations. We get analytical expressions for diffusion coefficients from which general relations between different diffusion modes at various temperatures are derived. Effective  diffusion coefficients are represented by prefactor and activation energy parameters. In general both these parameters show temperature dependence which is in contrast with the global Arrhenius behaviour of the diffusion known for simple systems.
 Such temperature dependence of  diffusion could explain some discrepancies in results obtained   from different experiments \cite{chvoj,chvoj2,bietti}.  
 Analysis of diffusion coefficients also shows that whereas dimer diffusion on Ag surface happens via translation together with rotation, on Cu surface both moves are separated, and fast rotation is limited only to cells, whereas slow intercell jumps decide about rate of translational diffusion of dimers.

\section{Calculation of diffusion coefficients}
Variational approach to the calculation of the diffusion coefficient from a set of master equations was proposed in Ref. \cite{Gortel1}. Since then it has been applied to many different systems of adatoms jumping on a crystal surface. So far mostly strictly theoretical systems have been considered. However, recently we have succesfully applied the variational approach to the diffusion of a single Ga adatom on the GaAs(001) surface c(4x4) $\alpha$ and $\beta$ reconstructions \cite{M;Z-K1} using the energy landscapes from Refs. \cite{Roehl1} and \cite{Roehl2}. It is also possible to solve a problem of correlated particles using our approach. Till now single adatom has always been treated as a simple particle, that is a particle whose state could be completely described by giving its position on the lattice. It could jump to one of the neighbouring sites on the lattice provided it wasn't already occupied. Such an adatom didn't possess any additional degrees of freedom like for example orientation or shape.

In this paper we describe dimer diffusion, namely Cu dimer on Cu(111) and Ag(111) surfaces. A dimer could be thought of as a compound particle or a cluster, that is an ensemble consisting of a certain number (two in the case of dimer) of simple particles that are bound together. Therefore, the move of the dimer should always be considered as the move of the whole, not as that of the individual constituents. The state of the dimer can be described fully by giving the position of its centre of mass, its length (the distance between the constituents) and its orientation along the underlying lattice. Therefore, the approach can be generalised to compound particles without any modifications.

Diffusion coefficient of a single particle jumping on a crystal lattice is defined as the mean square displacement over time \cite{gomer,tringides,ala,haus}
\begin{equation}
D_{nm}=\lim_{t\rightarrow\infty}\frac{1}{4t}<\Delta r_{n}(t)\Delta r_{m}(t)>.\label{diff_def}
\end{equation}
Indices n and m denote one of the space coordinates (x or y) on the lattice. $\Delta r_{n(m)}(t)$ is the displacement of the particle's centre of mass in the n(m) direction after time t. Since we follow the centre of mass, it doesn't matter if by a particle we mean a simple point particle or a cluster. Position of its centre of mass is always well defined. The above definition is used for extracting the diffusion coefficient from the data obtained from scanning tunneling microscopy or molecular dynamics methods.

In order to analyse diffusive motion on a crystal surface we can divide arbitrarily the lattice into unit cells which repeat themselves periodically over infinite surface. Origin of each cell has position $\vec{r}_{j}$ in the space and within given cell there are $m$ adsorption sites, whose locations relative to the cell's origin are given by $\vec{a}_{\alpha}$, where $\alpha=1,\ldots m$. Therefore, each adsorption site on the lattice is specified unambiguously by parameters (j,$\alpha$) and its position in the space is equal to $\vec{r}_{j}^{\alpha}=\vec{r}_{j}+\vec{a}_{\alpha}$. Physically, position of a given adsorption site corresponds to the position of the centre of mass of the particle (simple or compound) occupying that site. A particle occupying the site (j,$\alpha$) has the adsorption energy $E(\alpha)$. In general, it is possible for two or more different adsorption sites to have the same position. They will be distinguished by the index $\alpha$ though. For the coverage $\rho$ (ratio of adparticles at the surface to the total number of adsorption sites) the probability of finding a particle in the site of the energy $E_{\alpha}$ is equal to $P_{eq}(\alpha)=\rho\exp[-\beta E(\alpha)]/\sum_{\gamma}\exp[-\beta E(\gamma)]$. The summation is done over all the sites in the cell and $\beta=1/k_{B}T$ is the inverse temperature factor.

Diffusive motion on a surface consists of a series of thermally activated jumps between adsorption sites. According to the transition state theory, the jump rate (number of jumps per unit of time) is a function of the energy barrier along the path of the jump \cite{TST}. For a single jump from the (j,$\alpha$) site to the (l,$\gamma$) site it is equal to
\begin{equation}
W(j,\alpha;l,\gamma)=W_{\alpha,\gamma}=\nu\exp\{-\beta[\tilde{E}(j,\alpha;l,\gamma)-E(\alpha)]\},\label{jump_rate}
\end{equation}
$E(\alpha)$ is the adsorption energy of the initial site of the jump and $\tilde{E}(j,\alpha;l,\gamma)$ is the energy at the saddle point between the two sites involved in the jump. These are the energies that are found by means of ab-initio calculations \cite{Hayat1}. Variable $\nu$ is the diffusion prefactor, whose exact value is a subject for a separate study, however, its value is very often assumed to be of order of $10^{13}/s$. It  is sometimes also called the attempt frequency and its physical meaning is the number of attempts of the particle to jump out of its site per unit of time. The exponential part in (\ref{jump_rate}), on the other hand, is the probability of a successful attempt.
Time evolution of a system  is described by the Master equation
\begin{equation}
\frac{d}{dt}P(j,\alpha;t)=\sum_{l,\gamma}[W_{\gamma,\alpha}P(l,\gamma;t)-W_{\alpha,\gamma}P(j,\alpha;t)].\label{master_eq}
\end{equation}
$W_{\gamma,\alpha}$ are jump rates defined by (\ref{jump_rate}) and P(j,$\alpha$;t) is the probability that the site (j,$\alpha$) is occupied. We are looking for the diffusive eigenvector of Eqs (\ref{master_eq}).
We will calculate this eigenvector by the variational method. Details of this method are   described in Appendix A.

\section{Diffusion of $Cu$ monomer and dimer on $Cu(111)$ and $Ag(111)$ surfaces}
\begin{figure}
\begin{center}
\includegraphics[width=6cm]{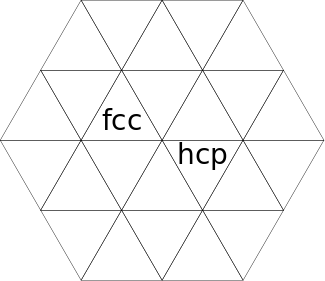}
\end{center}
\caption{Fcc(111) lattice with two possible energy minima for a Cu monomer.}
\label{fcc111}
\end{figure}

Both copper and silver crystallize in the face centered cubic structure and their (111) lattices consist of two types of sites called fcc and hcp, and arranged in the way shown in Fig. \ref{fcc111}. The lattice constant for silver is equal to $4.09\AA$ compared to $3.61\AA$ for copper. This difference in lattice constant is due to the interaction strength between substrate atoms and it also determines the shape of the energetic surface that is seen by adsorbed Cu atoms.

The energy landscape for Cu monomer and Cu dimer on Cu(111) was calculated in Ref. \cite{Marinica1} and on Ag(111) in Ref. \cite{Hayat1} using ab-initio methods. On the base of these calculations we can see that Ag surface observed by Cu adatom in hetero-diffusion process is more smooth. There is no evident difference between jump rates inside and outside surface cell. At the same time in the homo-diffusion process Cu/Cu(111) jumps from one cell to the neighboring one are the slowest ones and determine the global diffusion rate.
We use here the potential energy surfaces from Refs. \cite{Marinica1,Hayat1} to calculate the diffusion coefficient for both Cu monomer and dimer. For the dimer we calculate also rotational diffusion coefficient and compare it with the translational one.

\begin{figure}
\begin{center}
\includegraphics[width=7cm]{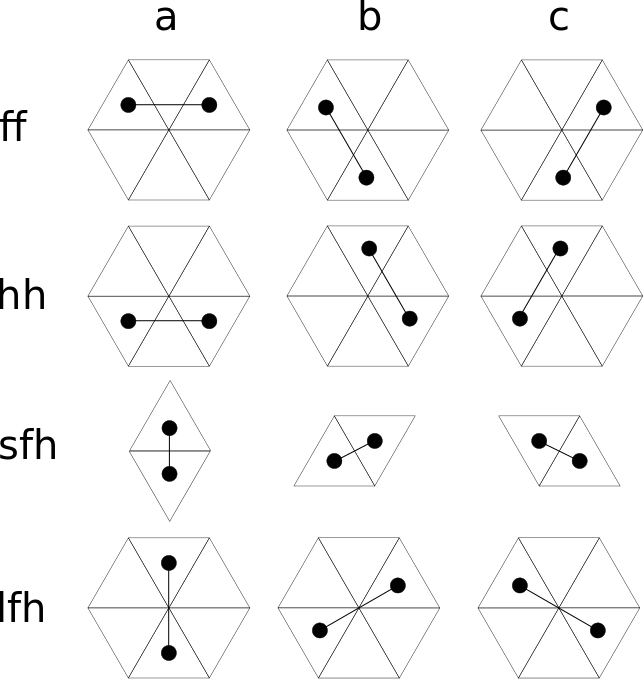}
\end{center}
\caption{Possible configurations of Cu dimer on the fcc(111) surface. The sfh configurations appear only on the Ag(111).}
\label{conf}
\end{figure}

A single adatom can occupy either fcc or hcp site. These sites differ from each other by the position of the atoms lying directly below.  An adatom occupying an fcc site will therefore form bonds of different lengths and directions with underlying surface atoms than the one that occupies an hcp site. It means that the potential energy of the adatom will depend on whether it resides at an fcc or an hcp site. However, according to both the experiment \cite{Morgenstern1,Morgenstern2,Repp1} and the calculations \cite{Marinica1,Hayat1}, the  energy difference of these two sites is very small, of order of single meV at  both  Cu(111) and Ag(111) surfaces.

At the same time  dimers  formed at these surfaces see completely different landscape of potential  energy. It is observed that  dimers  become  quite stable and long-lived  structures \cite{Morgenstern1,Morgenstern2} and it is possible to study their diffusional movement. Dissociation of the Cu dimer on Cu(111) has been never observed experimentally or in simulations. According to \cite{Marinica1} the energy barrier for such a process is 450 meV, which is much higher than for most of the possible jumps of the dimer. On the other hand, dissociation of the Cu dimer on Ag(111) has occured at 700 K in molecular dynamics simulations \cite{Hayat1}, which suggests that the corresponding energy barrier is reduced on that surface compared to Cu(111). However, the dissociation was only temporary and after a few ps the constituents recombined forming the dimer again. Therefore, the dimers can be treated as stable particles in our temperature range.

 According to Ref. \cite{Hayat1}  Cu dimer on Ag(111) can be found in one of the four possible energetic minima (see Fig. \ref{conf}). They differ by the type of site occupied by each single particle in the dimer (fcc or hcp) and by the distance between those particles. Different dimer configurations are called ff (two fcc sites), hh (two hcp sites), sfh (short fh) and lfh (long fh). The ff and hh states differ only by the type of occupied sites while the distance between the particles is the same in both cases. On the other hand, in sfh and lfh states, particles occupy the same pairs of  sites but the distance between them can be either short (sfh) or long (lfh). At Cu(111) surface  Cu dimer can be found in ff, hh or lfh configurations only. The sfh state is not built on this lattice \cite{Marinica1}. Each of the dimer's energy minima  on this surface can appear in three variants which differ by their relative angular orientation. All the possible configurations of the Cu dimer on the fcc(111) lattice are shown in Fig. \ref{conf}.

It has been argued \cite{Hayat1} whether the sfh and lfh states are actually potential minima on the Ag(111) lattice or just metastable states. They lie only a few meV below the calculated transition states to the ff and hh states and the potential energy landscape around sfh and lfh states is almost flat. However, authors treat these configurations as separate states and in our calculations below we shall also consider the sfh and lfh sites as energy minima. In other situation all barriers for jumps between states should be calculated from the beginning and then they can be used in our formula giving similar results as the one shown below.
\begin{figure}
\begin{center}
\includegraphics[angle=-90,width=7cm]{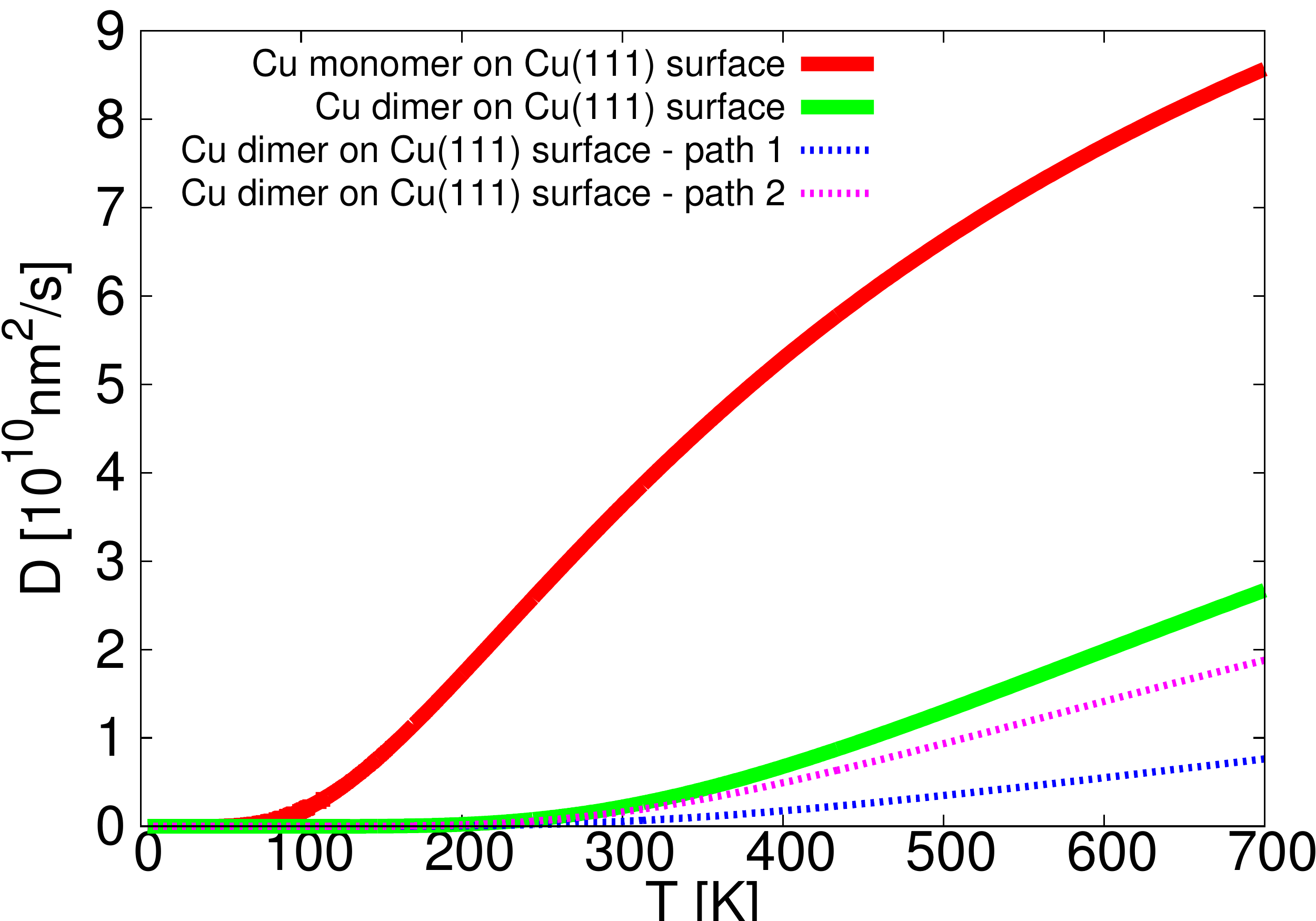}
\includegraphics[angle=-90,width=7cm]{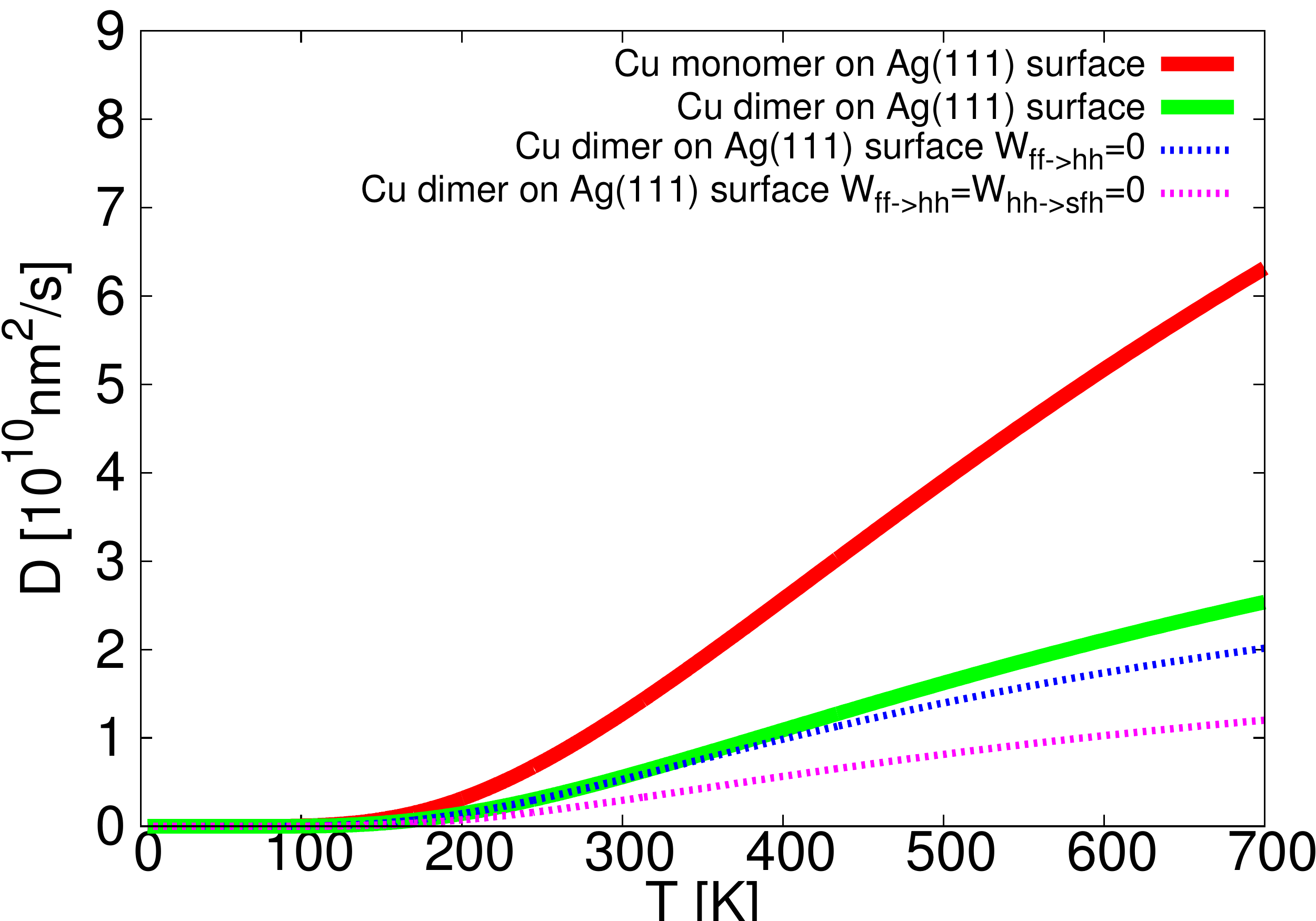}
\end{center}
\caption{Diffusion coefficient for Cu monomer and  Cu dimer on the Cu(111) (top) and Ag(111) (bottom) surfaces as the function of the temperature. $\nu=10^{13}/s$.}
\label{diff_coeff}
\end{figure}

\begin{figure}
\begin{center}
\includegraphics[angle=-90,width=7cm]{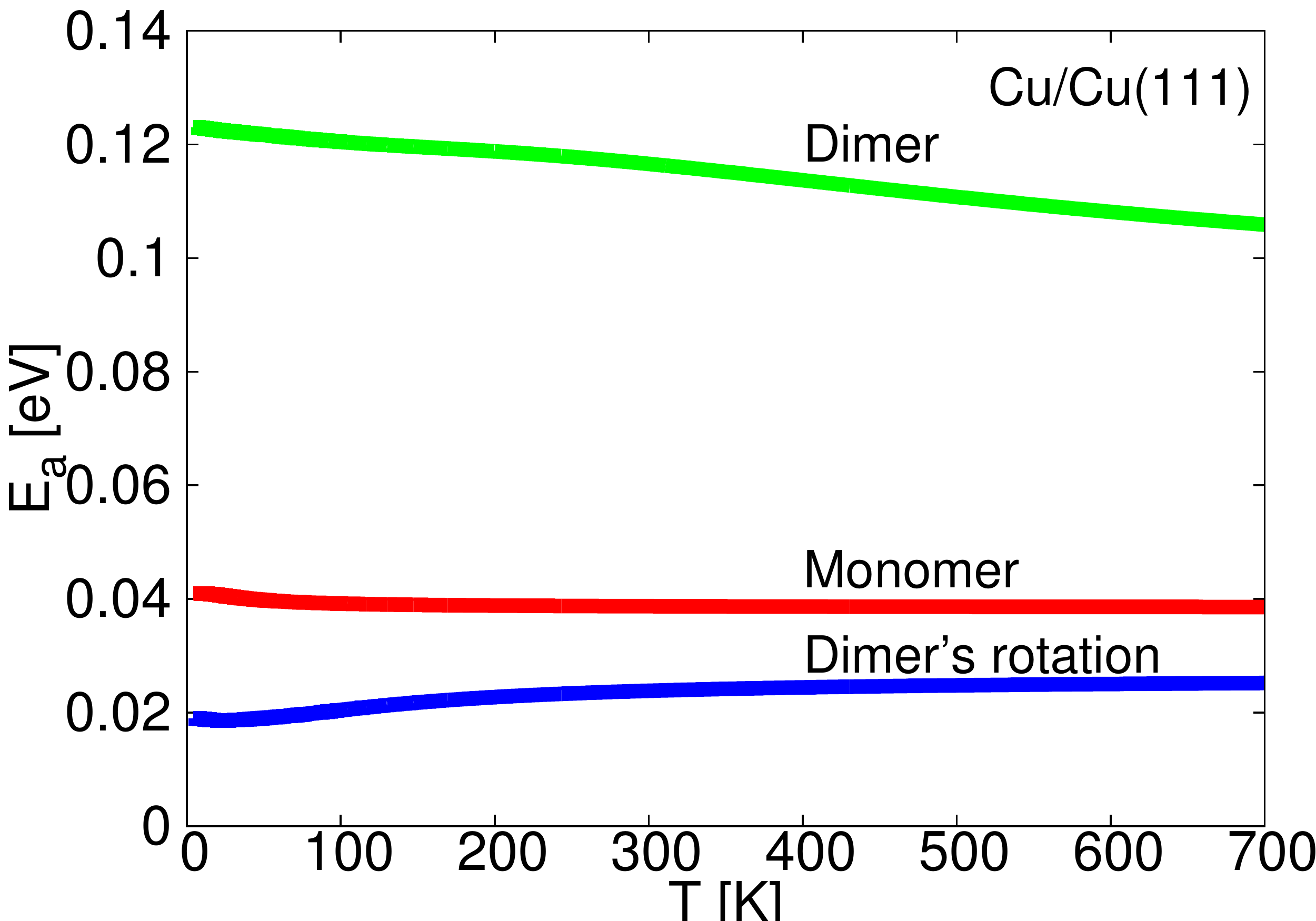}
\includegraphics[angle=-90,width=7cm]{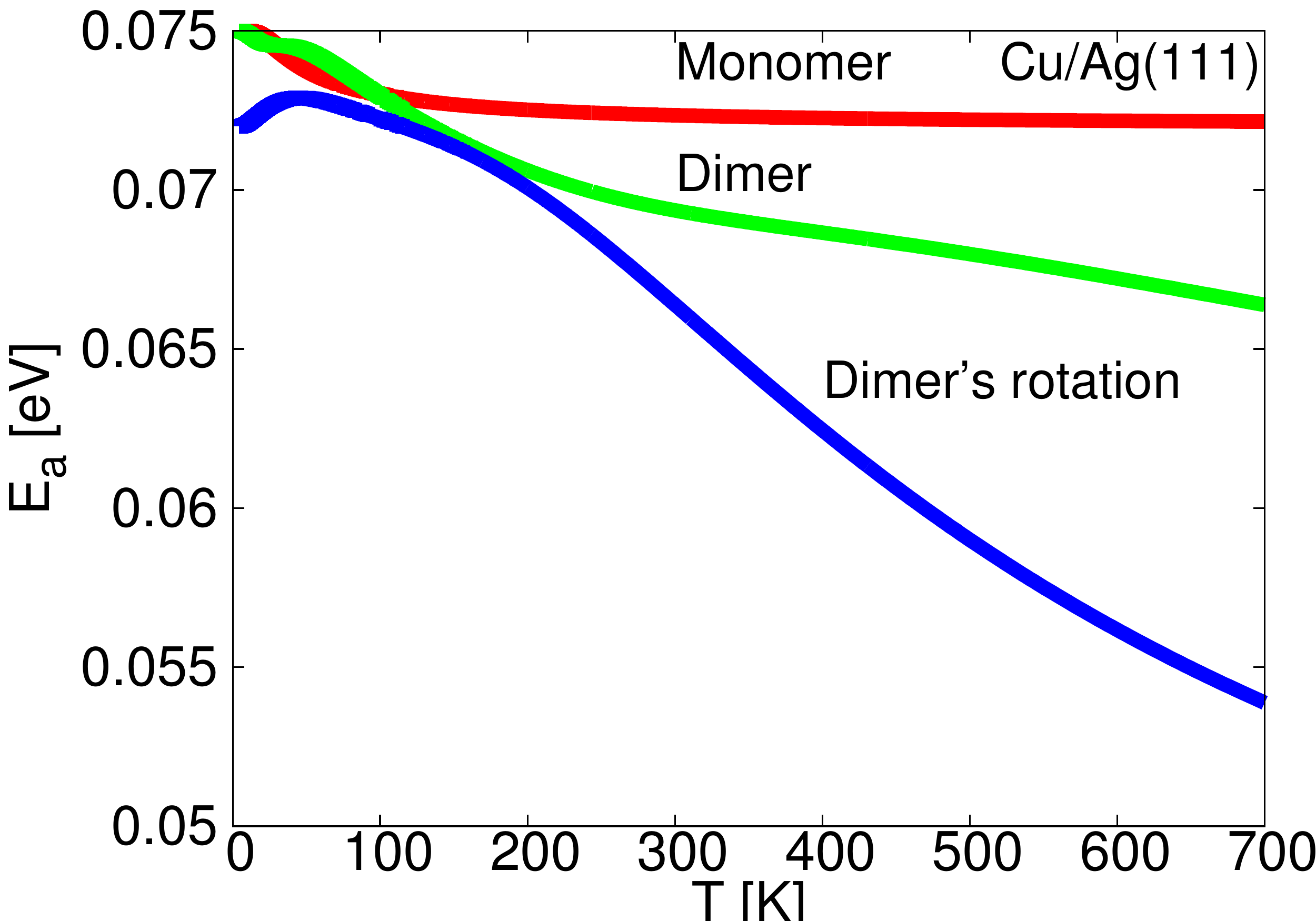}
\end{center}
\caption{Effective activation energy for the diffusive motion of Cu monomer and dimer, and for the rotational diffusive motion of dimer on the Cu(111) (top) and Ag(111) (bottom) surfaces as the function of the temperature.}
\label{eff_act_en}
\end{figure}
\subsection{Monomer}
Structure of the energy landscape as seen by Cu monomer is the same for both  Cu(111) and  Ag(111) surfaces. Both the energetic minima and the barriers are found at the same positions. The only difference is the exact values of the energies. Therefore, we can derive a general formula for the monomer and then substitute specific values for both  Cu(111) and the Ag(111) surfaces in order to calculate the diffusion coefficients.

As it is apparent from the first-principles calculations \cite{Marinica1,Hayat1} and from the experiments \cite{Morgenstern1,Morgenstern2,Repp1}, Cu monomers are slightly more stable at fcc sites than at hcp sites. For Cu(111) surface the difference between energies at those two sites is equal to 5 meV \cite{Marinica1} and for the Ag(111) lattice it is equal to 6 meV \cite{Hayat1}. The energetic barrier for the jump from an fcc to an hcp site on Cu is equal to 0.041 eV and to 0.075 eV on Ag. The barriers for inverse processes are 0.036 eV and 0.069 eV, respectively. We use these data in order to calculate the diffusion coefficient for a single Cu adatom on the Cu(111) and the Ag(111) surfaces. The variational parameters in (\ref{var_form}) are equal to $\vec{0}$ because of the system symmetry. Therefore, we just have a sum of two terms, each related to one type of jump, which can be also expressed as
\begin{equation}
D=\frac{a^2}{2}\frac{W_{fcc\rightarrow hcp}W_{hcp\rightarrow fcc}}{W_{fcc\rightarrow hcp}+W_{hcp\rightarrow fcc}}\label{diff_mon}
\end{equation}
The parameter $a$ in the above formula (and in all  equations in this article) is the length of the lattice constant of the underlying surface, which is also equal to the bond length of Cu or Ag. It is related to the bulk lattice constant $A$ by $a=\frac{A\sqrt{2}}{2}$ what gives $a_{Cu}=2.55 \AA$ for Cu and $a_{Ag}=2.89 \AA$ for Ag surface.

According to (\ref{diff_mon}), the diffusion of Cu monomer on Cu(111) and Ag(111) is given by the same formula as this for  the one-dimensional diffusion of a single particle in the potential with two alternating minima. The temperature dependence of the monomer's diffusion coefficients is shown in Fig.~\ref{diff_coeff} for Cu surface in top panel and for Ag surface in bottom panel, in both cases plotted as top lines. It can be seen that monomer diffusion on Ag surface is slightly slower than this on Cu lattice. Assuming a local Arrhenius behavior of the diffusion coefficient $D=D_{0}\exp(-\beta E_{a})$ it is possible to calculate the effective activation energy $E_a=-\frac{\partial\ln{D}}{\partial\beta}$. For the monomer it is equal to
\begin{equation}
E_{a}=\frac{W_{fcc\rightarrow hcp}\Delta E_{hcp\rightarrow fcc}+W_{hcp\rightarrow fcc}\Delta E_{fcc\rightarrow hcp}}{W_{fcc\rightarrow hcp}+W_{hcp\rightarrow fcc}}
\end{equation}
It is obvious that a quantity defined in such a way depends on the temperature. $E_a$ for monomer on Cu lattice is shown in top panel of Fig.~\ref{eff_act_en}, drawn with the middle line and $E_a$ for monomer on Ag lattice it is shown with the top line in the  bottom panel of Fig.~\ref{eff_act_en}. As seen in both cases $E_a$ does not change much, it starts at lower temperatures from the value equal to the height of the larger of the two barriers and at higher temperatures it goes down approaching the value which is the average of both barriers. Therefore, for Cu(111) surface the effective activation energy goes from  41 meV to 39 meV. We can compare then with low temperature  experimental value 37$\pm$5 \cite{Repp1}. For Ag(111) surface activation energy starts at 75 meV at low temperatures compared to 65$\pm$9 meV in the experiment \cite{Morgenstern1}, then it  goes down reaching 72 meV at high temperatures.

\begin{figure}
\begin{center}
\includegraphics[angle=-90,width=7cm]{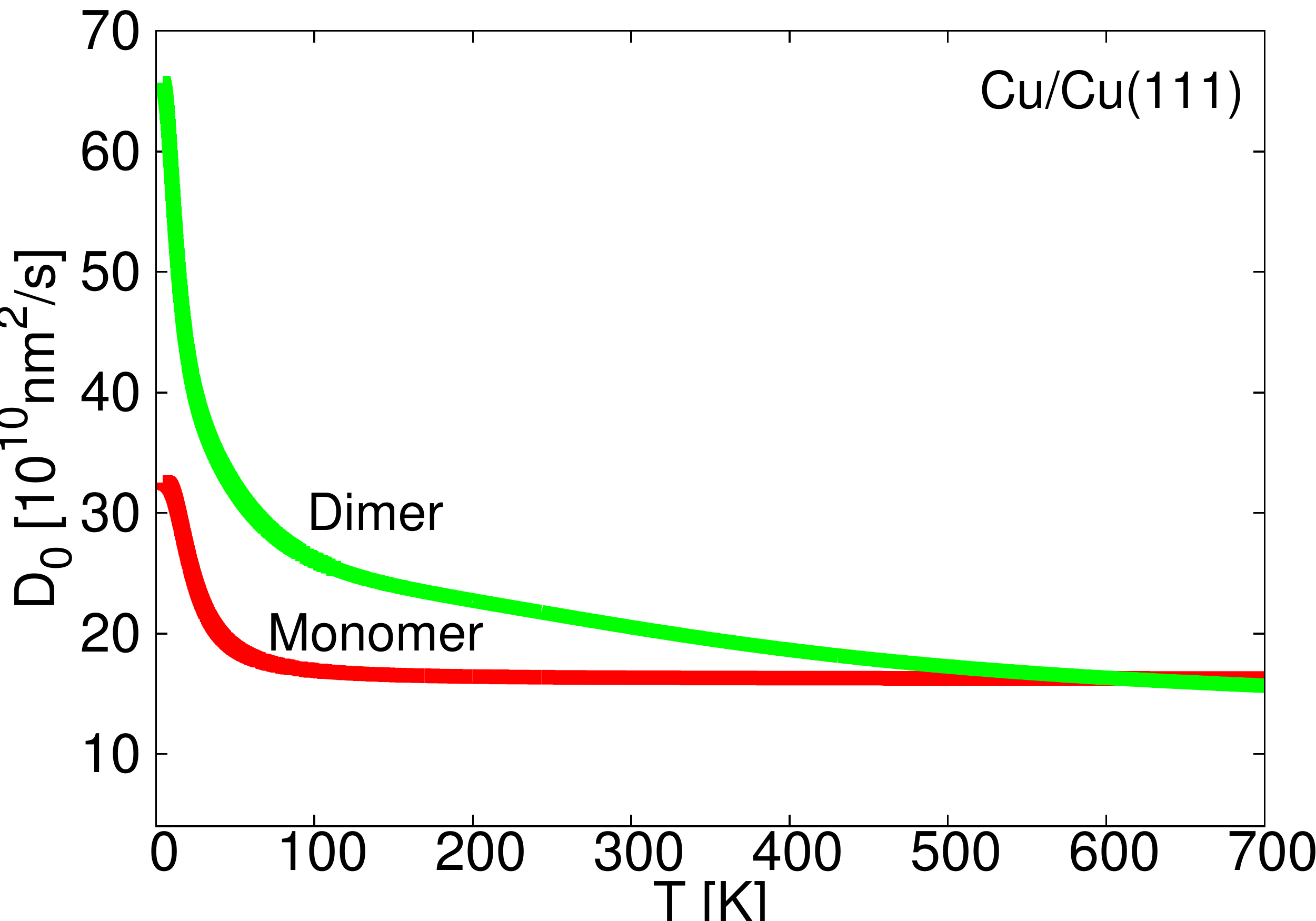}
\includegraphics[angle=-90,width=7cm]{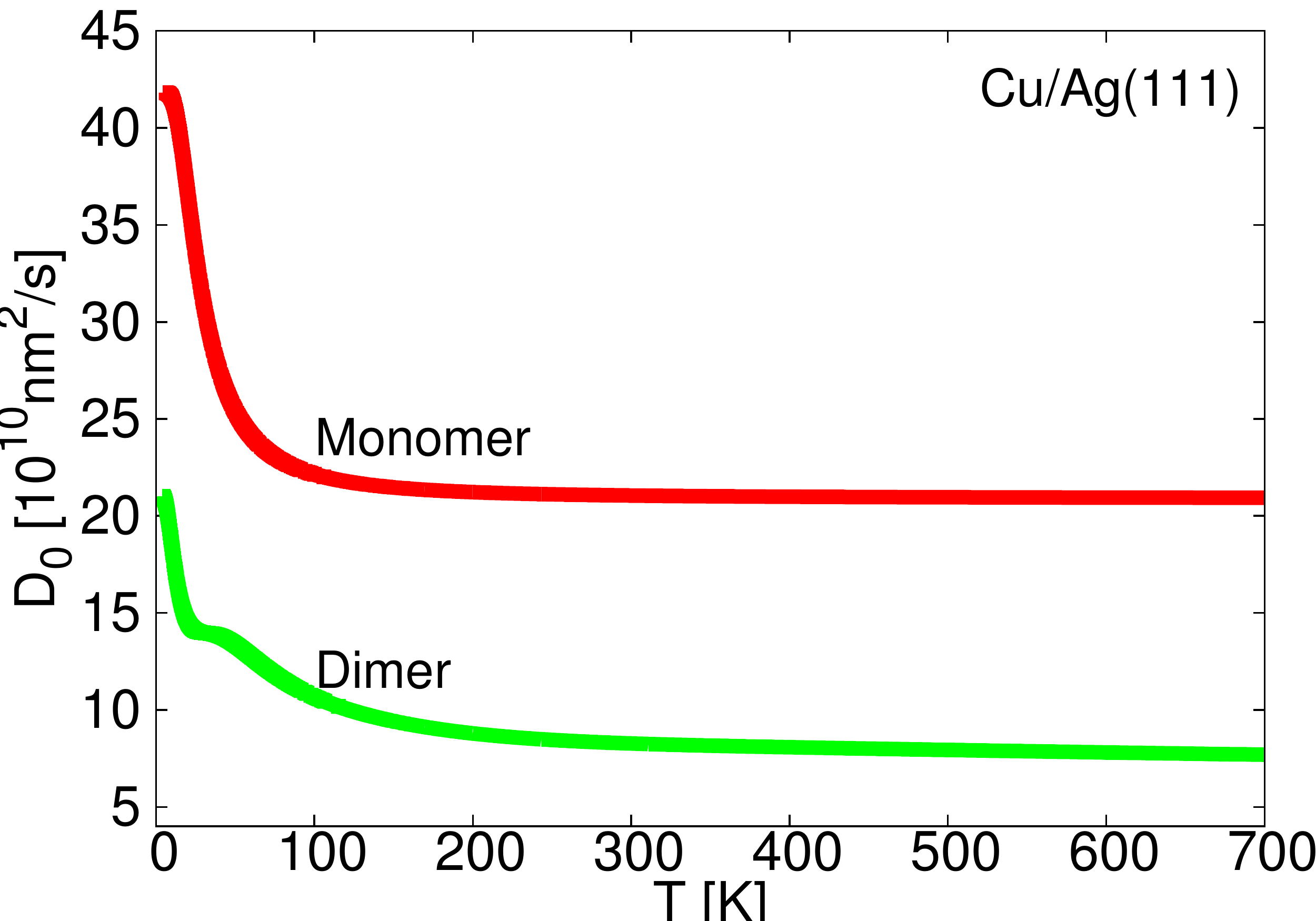}
\end{center}
\caption{Effective prefactor for the diffusive motion of Cu monomer and dimer on the Cu(111) (top) and Ag(111) (bottom) surfaces as the function of the temperature. $\nu=10^{13}/s$}
\label{eff_pref}
\end{figure}

Knowing the effective activation energy, it is also possible to calculate the effective prefactor $D_{0}=D\exp(\beta E_{a})$. Its behaviour is shown in Fig. \ref{eff_pref} with the  bottom line in the top panel and with the  top line in bottom panel. In both cases at low temperatures they decrease rapidly  up to  about 100 K and then remain constant at higher temperatures.

\begin{figure}
\begin{center}
\includegraphics[width=9cm]{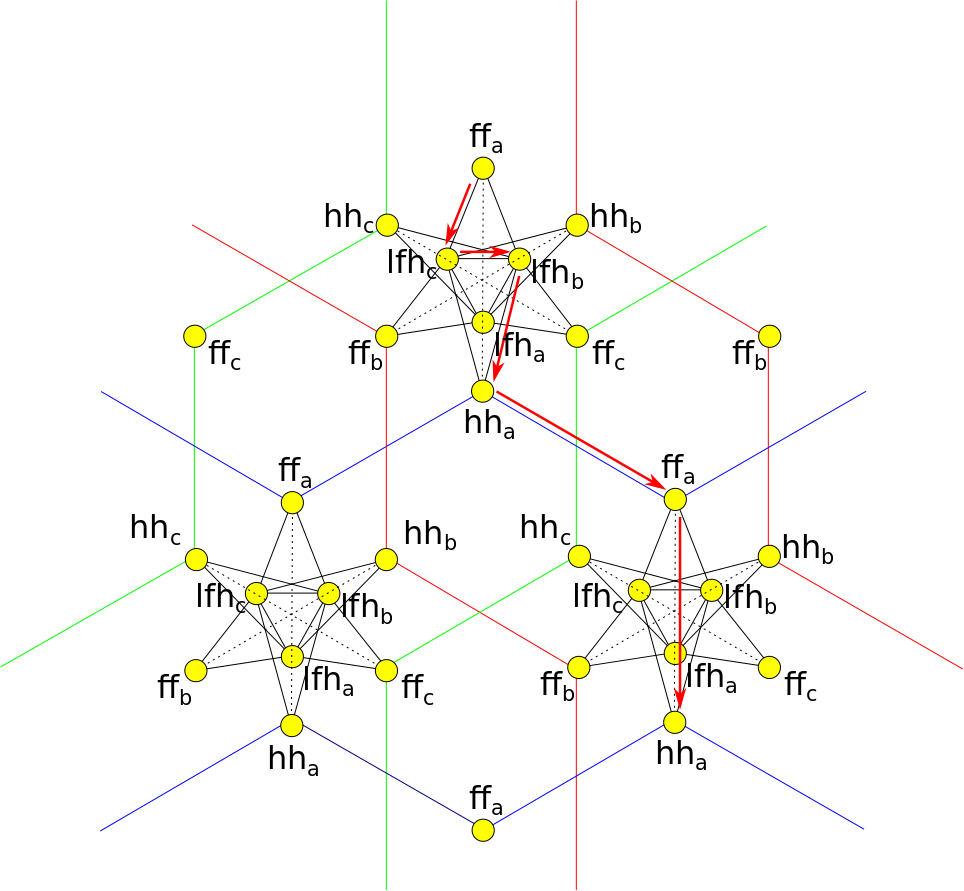}
\includegraphics[width=6cm]{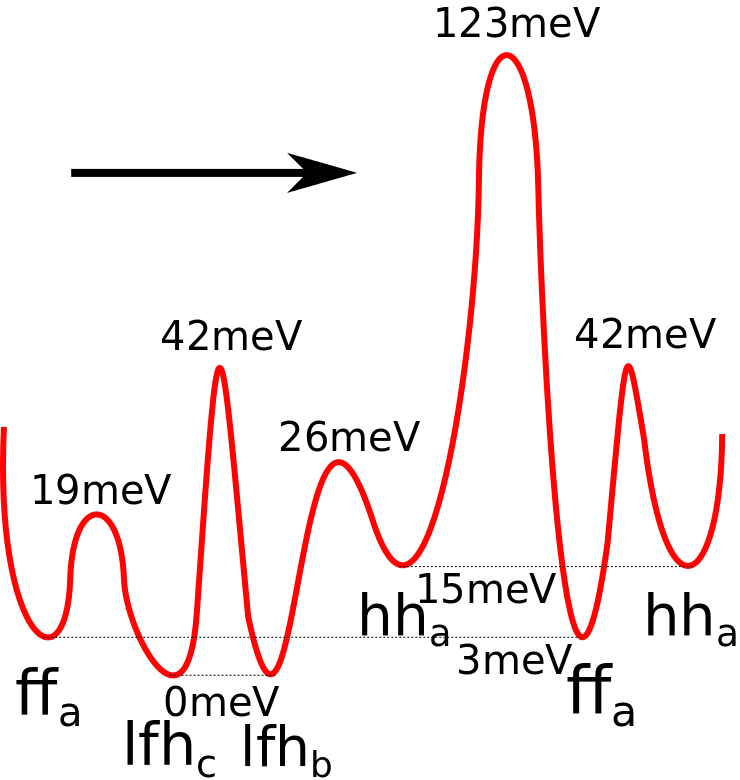}
\caption{Scheme of the transitions on the Cu(111) (top) surface  and energy  profile (bottom) along the marked path for a single Cu dimer. Names of the sites are assigned according to Fig.~\ref{conf}.}
\label{jumps}
\end{center}
\end{figure}
\begin{figure}
\begin{center}
\includegraphics[width=7cm]{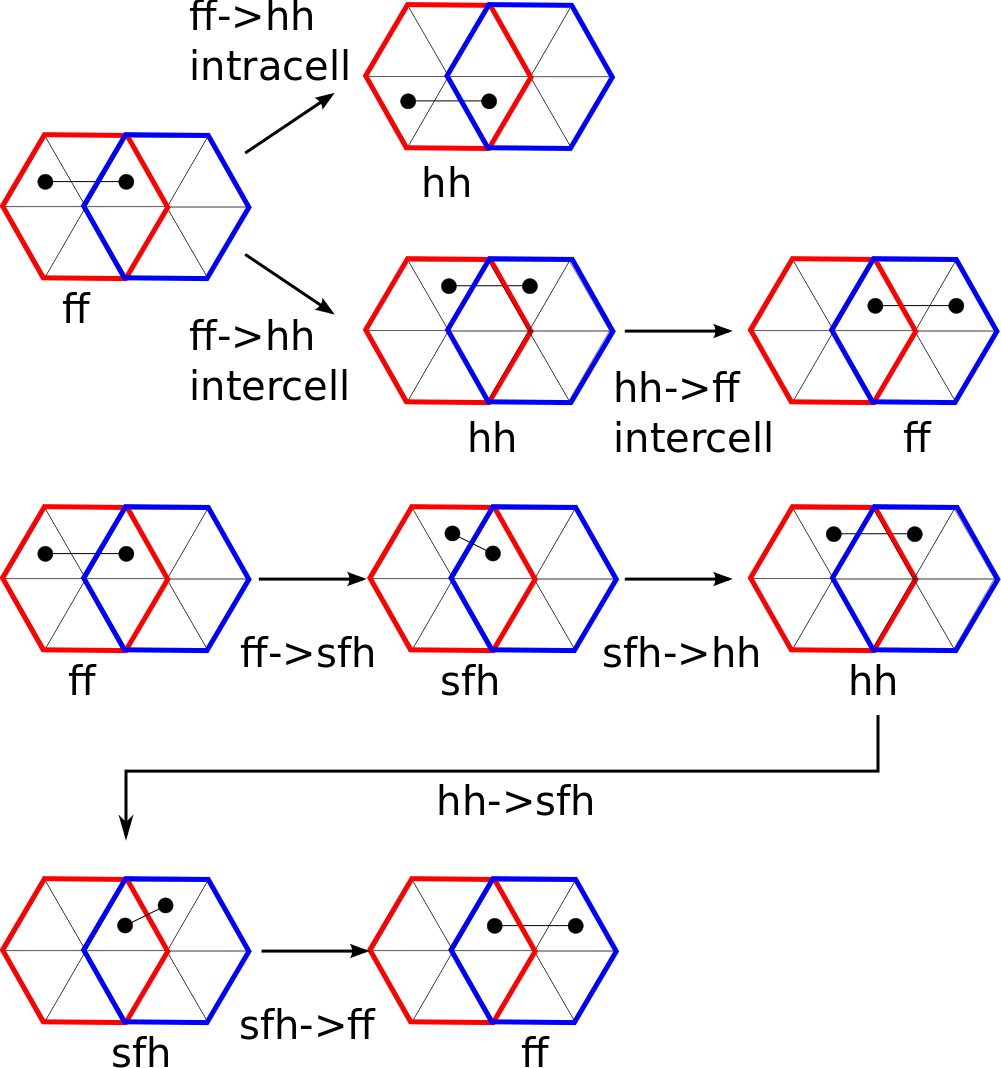}
\caption{Realizations of intercell jumps via  hh$\rightarrow$ff jumps at Cu(111) surface (top) and via sfh state at Ag(111) surface (bottom). \label{intercell}}
\end{center}
\end{figure}
\begin{figure}
\begin{center}
\includegraphics[width=14cm]{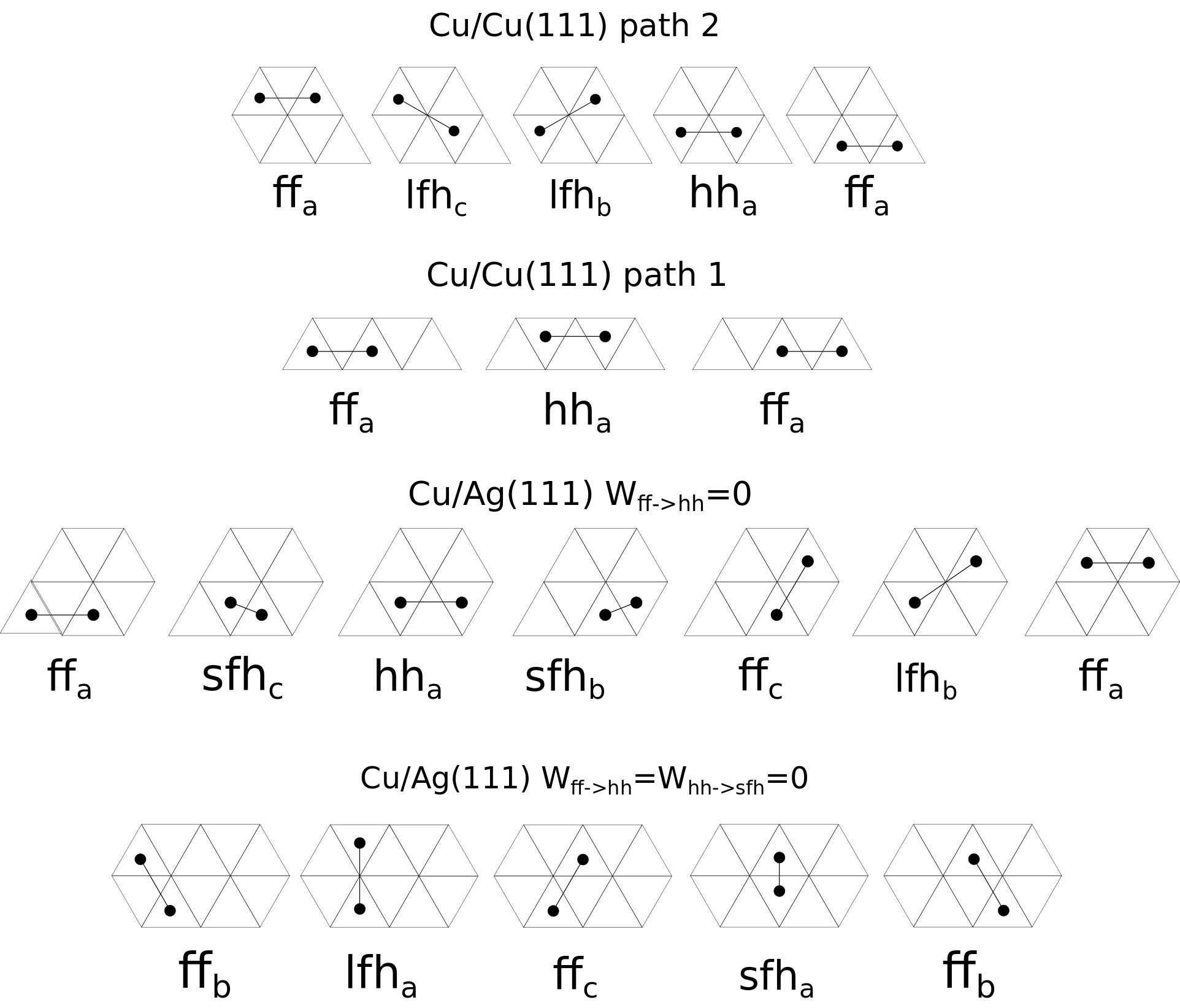}
\caption{Different diffusion modes on Cu(111) (top) and Ag(111) (bottom) surfaces. \label{path_dimer}}
\end{center}
\end{figure}
\begin{figure}
\begin{center}
\includegraphics[angle=-90,width=8cm]{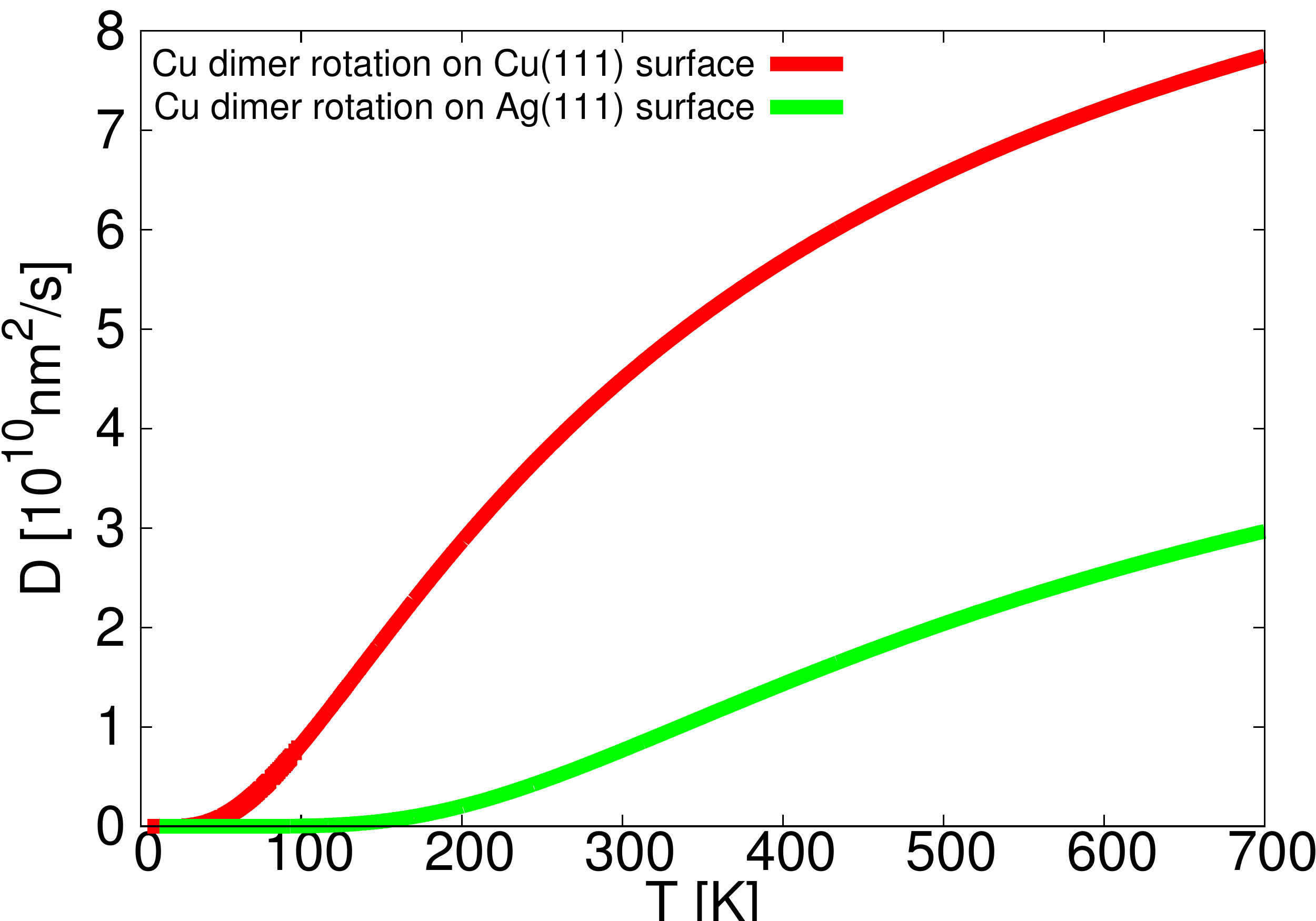}
\end{center}
\caption{Rotational diffusion coefficient for  Cu dimer on the Cu(111)  (top) and  Ag(111) (bottom) surfaces as the function of the temperature. $\nu=10^{13}/s$}
\label{diff_rot_coeff}
\end{figure}
\begin{figure}
\begin{center}
\includegraphics[angle=-90,width=8cm]{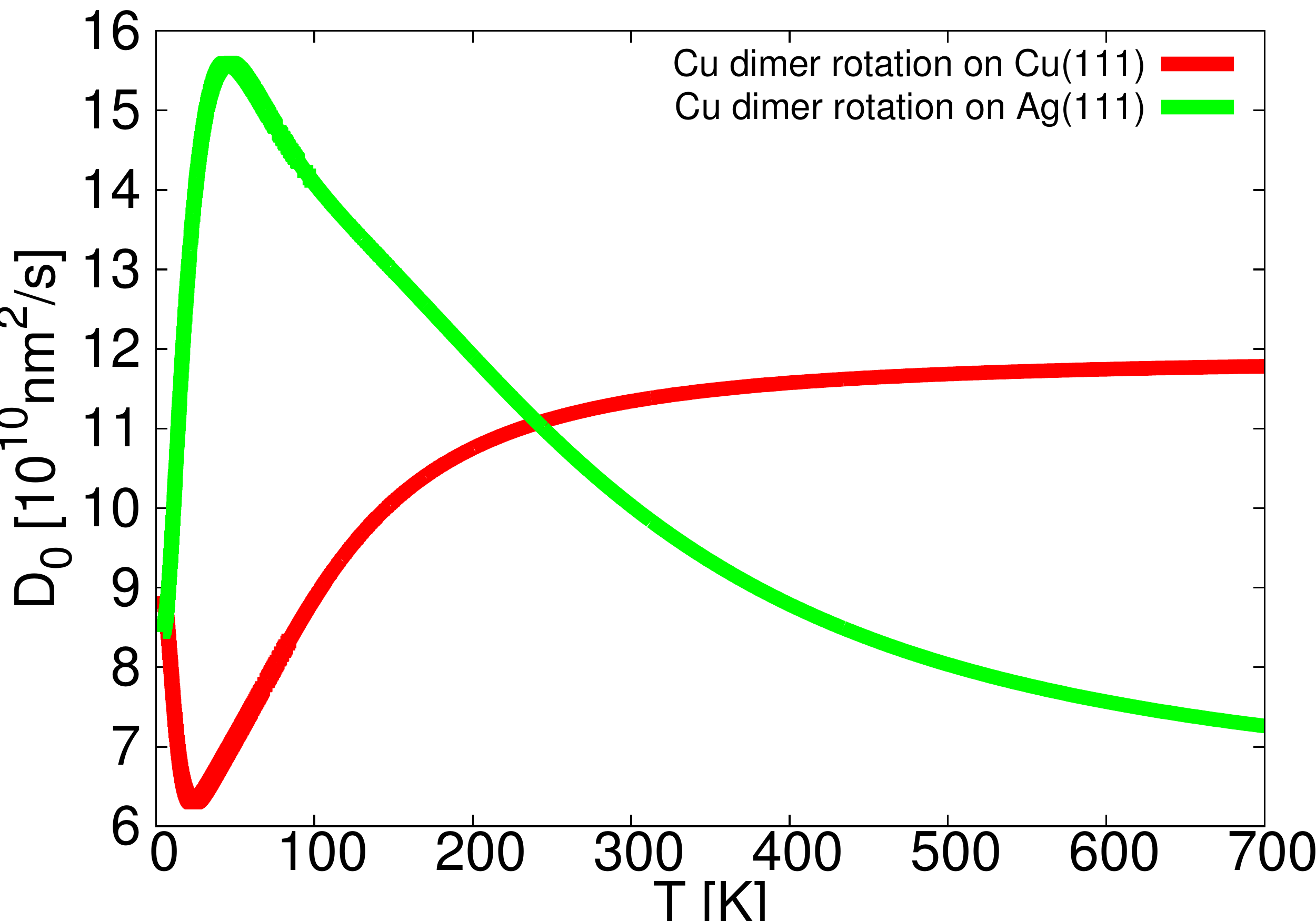}
\end{center}
\caption{Effective prefactor for the rotational diffusive motion of dimer on the Cu(111) (top) and Ag(111) (bottom) lattices as the function of the temperature. $\nu=10^{13}/s$.}
\label{eff_pref_rot}
\end{figure}
\begin{figure}
\begin{center}
\includegraphics[width=9cm]{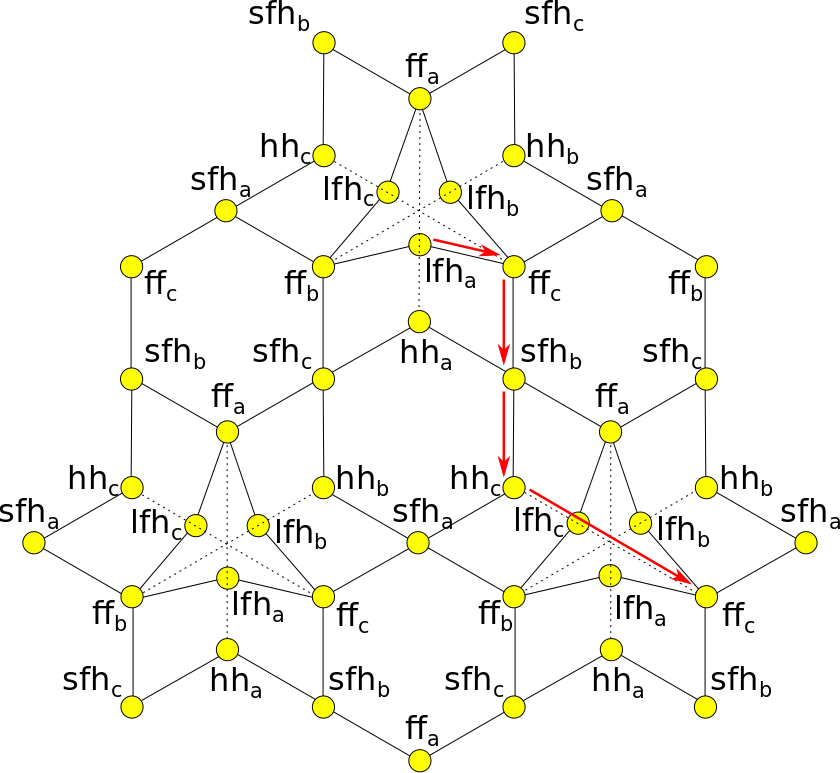}
\includegraphics[width=6cm]{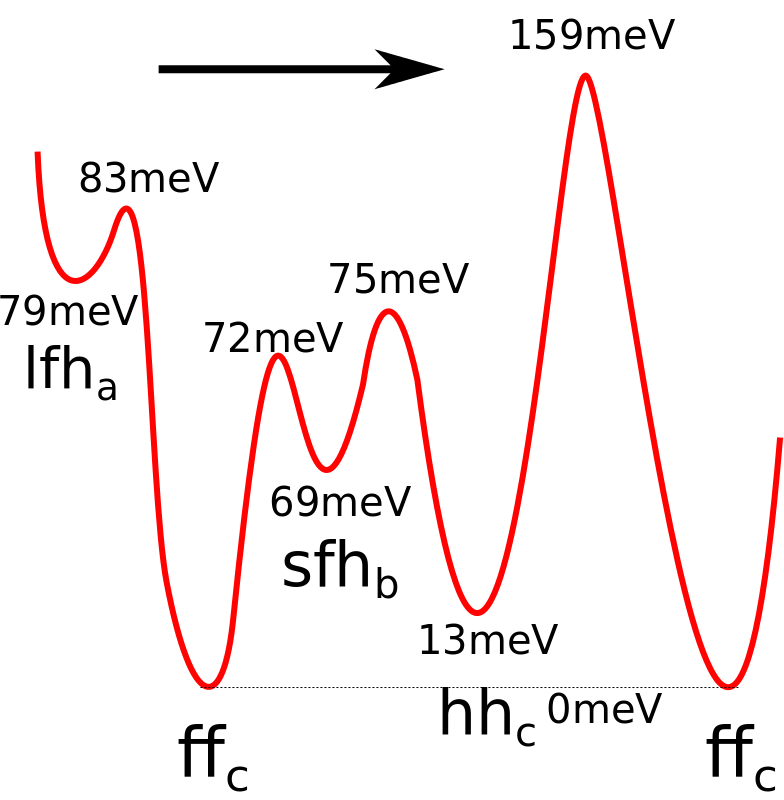}
\end{center}
\caption{Scheme of the transitions on the Ag(111) (top) surface and energy profile (bottom) along the marked path for a single Cu dimer. Names of the sites are assigned according to Fig.~\ref{conf}.}
\label{jumps2}
\end{figure}

\subsection{Dimer diffusion}
For Cu dimer on the fcc(111) surfaces the situation is more complex than for the monomer. As discussed above we have the following dimer configurations: ff, hh, fh that differ by the positions of two atoms of dimer. As we shall see below the distance between the atoms also matters and in general we will have sfh (short fh) and lfh (long fh) positions. Moreover each configuration occurs in three different orientations. All the possible configurations found on the fcc(111) lattices are shown in Fig. \ref{conf}. The map of the possible transitions between all those sites is quite complex and differs significantly between Cu and Ag surfaces, therefore, below we shall discuss both cases separately. The energy landscape and values of energy for both the sites and the barriers  have been taken from Ref. \cite{Marinica1} (for Cu(111)) and from Ref. \cite{Hayat1} (for Ag(111)) and used below to calculate the diffusion coefficients of Cu dimers on those lattices.

\subsubsection{Cu dimer on Cu(111) lattice}
Only three  types of equilibrium  states from Fig. \ref{conf} are found for  dimer on Cu(111) lattice \cite{Repp1,Marinica1}. These are ff, hh and lfh. Configuration sfh has too high energy at this lattice hence it cannot be used by dimer to realize intercell jumps. Energies of dimer states and barriers for  transitions between those states has been worked out in Ref. \cite{Marinica1}. We plotted the map of all the possible jumps in our system  in  Fig. \ref{jumps}. Yellow circles represent adsorption sites for the dimer (its centre of mass on the lattice) and the lines represent transitions. Three lfh sites with different subscripts have actually the same position but have been drawn as separate ones for clarity. The black lines represent intracell jumps, which are limited to a finite area on the lattice. The dashed black line between ff and hh sites means that this jump is directly between these sites and does not pass through the lfh site. Blue, red and green lines represent intercell motion in different directions, that is the one which goes beyond given  cell. Energy landscape along the path marked in red is plotted in the lower part of Fig. \ref{jumps}. Plotted path leads dimer from one cell to the next one. Each cell is visible in top panel of Fig \ref{jumps} in the form of star. Intercell jumps have  much higher energy barrier than intracell ones. In our calculations we shall take into account all the intracell jumps and the lowest in energy intercell motion, namely the concerted sliding between ff and hh sites, which has an energetic barrier of 120 meV. It is shown in  Fig. \ref{intercell} how this particular intercell jump is realized and how it differs from the  intracell jump of the same type. Barriers for all the remaining intercell jumps (including the dimer's dissociation) are at least three times  larger, therefore these  jumps are  highly unlikely and will be omitted in further considerations. 

We denote rate of the intercell jump  by $V_{ff\rightarrow hh}$, while the intracell jumps will have  rates $W_{ff\rightarrow hh}$, $W_{ff\rightarrow lfh}$, $W_{hh\rightarrow lfh}$ and $W_{lfh\rightarrow lfh}$.  Symbols with tilde over them will be equal to corresponding rates multiplied by the equilibrium probability of the initial site's occupancy, for example $\tilde{V}_{ff\rightarrow hh}=P_{eq}^{ff}V_{ff\rightarrow hh}$. We shall use this notation in the whole article.

It can be seen in Fig. \ref{jumps} that the system possesses three axes of symmetry (one along the a sites, one along the b sites and one along the c sites). The lattice sites labeled by different subscripts (a, b or c) are energetically equivalent, that is they have exactly the same potential energy and the same neighboring sites with the same jump rates to them. In order to apply our variational method, we need to introduce some parameters that depend on the state of the system. The state of the system is given completely by specifying in which of the sites shown in  Fig. \ref{conf} the dimer currently resides. Each of those sites should have its own geometric vector phase, which can be written for convenience in the angular form $\vec{\delta}=\delta_{r}\left(\cos{\delta_{\phi}},\sin{\delta_{\phi}}\right)$. The form of these phases should reflect the symmetry of the potential felt by the dimer at given  site. For   site $a$ it can be seen that there is symmetry with respect to the y axix, therefore we should have $\delta_{\phi}^{a}=\frac{\pi}{2}$, which gives $\vec{\delta^{a}}=\left(0,\delta_{r}^{a}\right)$. The angular parts of the corresponding $b$ and $c$ phases should then be equal to $\frac{7\pi}{6}$ and $\frac{11\pi}{3}$, respectively. Therefore, the only variational parameters will be the radial parts, that is $\delta_{r}^{ff}$, $\delta_{r}^{hh}$ and $\delta_{r}^{lfh}$. From now on we will omit the subscript r.

We insert such defined phases into the variational formula (\ref{var_form}) and, using simple differentiation, minimise the expression with respect to the parameters. Resulting expression for phases are given in Appendix B. Finally the diffusion coefficient for Cu dimer at Cu(111) surface is
\begin{align}
D=&\frac{3}{4}a^{2}\tilde{V}_{ff\rightarrow hh} \left\{ 1+3\left[\tilde{W}_{ff\rightarrow hh}(\tilde{W}_{ff\rightarrow lfh}+\tilde{W}_{hh\rightarrow lfh}) \right. \right. \nonumber \\
& \left. + 2\tilde{W}_{ff\rightarrow lfh}\tilde{W}_{hh\rightarrow lfh}] \right] \left [(\tilde{W}_{ff\rightarrow hh}+2\tilde{V}_{ff\rightarrow hh}) \right.\nonumber \\
& \left.\left.(\tilde{W}_{ff\rightarrow lfh}+\tilde{W}_{hh\rightarrow lfh})+2\tilde{W}_{ff\rightarrow lfh}\tilde{W}_{hh\rightarrow lfh}\right] ^{-1} \right \}.
\label{11}
\end{align}
As expected, the intercell jump rate $\tilde{V}_{ff\rightarrow hh}$ is fundamental for the diffusive motion of the dimer. The first term in the above sum describes the diffusion controlled exclusively by the $ff\rightarrow hh$ intercell jumps, which are quite slow. However, the second term has a large contribution to the total value of the diffusion coefficient. It describes the motion which is partially realised by the much faster intracell jumps. Therefore, we see that even though those jumps are unable to trigger the diffusion by themselves, they can enhance it significantly by bypassing some of the slow intercell jumps. These two modes of diffusion: one omitting  intracell jumps given by first term of sum (\ref{11}) and the second through the inter and intracell jumps  are illustrated in top panel of Fig.~\ref{path_dimer}. Fig. \ref{diff_coeff} shows diffusion coefficients for both modes. It is seen that the  one plotted with  the bottom line is much slower than the second one what means that appropriate diffusion path is less probable.

It should be also noted that jump rate $\tilde{W}_{lfh\rightarrow lfh}$ is not included in the above expression. It is understandable, since those jumps only rotate dimer between its different angular orientations. They do not move the dimer centre of mass at all. However, it is also possible to calculate the rotational diffusion coefficient. By analogy to (\ref{diff_def}) it is defined as
\begin{equation}
D_{r}=\lim_{t\rightarrow\infty}\frac{1}{2t}<\theta^2>.
\end{equation}
The variational formula (\ref{var_form}) can be used in the same form, if instead of the translational degree of freedom $\vec{r}$ we use the angular one $\theta$. The variational phases coupled to the angle $\theta$ go to zero, and it is easy to understand as all the sites are symmetric with respect to change of the angle. Thus the rotational diffusion coefficient for the Cu dimer on the Cu(111) is equal to
\begin{equation}
D_{r}=\frac{\pi^{2}}{18}(\tilde{W}_{ff\rightarrow lfh}+\tilde{W}_{hh\rightarrow lfh}+2\tilde{W}_{lfh\rightarrow lfh}).
\end{equation}
It implies that the rotational diffusion is realised in three independent ways. The contribution of the $lfh\rightarrow lfh$ jump is twice as large as these of $ff\rightarrow lfh$ and $hh\rightarrow lfh$ because there is only one $lfh\rightarrow lfh$ jump needed in order to rotate the dimer by 120 degrees. Both $ff\rightarrow lfh$ and $hh\rightarrow lfh$ need to be performed twice for the same effect.

It can be seen in Fig. \ref{diff_coeff} that the dimer diffusion coefficient is much smaller than that for the monomer. It is more difficult to compare  translational and rotational motion of dimer because one is measured in $nm^2/s$ and the second in $rad^2/s$. In order to compare them both we multiplied $ D_r$ by $d^2/4$, square of half of the dimer length. Such value can be attributed  to the diffusion of one of the atoms of dimer in the rotational movement. In Fig. \ref{diff_rot_coeff} the upper  line shows Cu dimer rotation in this case and it can be seen that it is as fast as monomer diffusion over the surface and at the same time much faster than dimer translational diffusion. It means that dimer rotates at given cell and this rotation does not couple to the translational diffusion. We also calculated the activation energies for these motions just as we have already done for the monomer. The formula for the translational dimer's motion is too complex to write it down, therefore we just plot the result in Fig. \ref{eff_act_en}. The expression for activation energy for the dimer's rotation is
\begin{align}
E_{a}&=\left[ (E_{ff}+\Delta E_{ff\rightarrow lfh})\tilde{W}_{ff\rightarrow lfh}+(E_{hh}+\Delta E_{hh\rightarrow lfh}) \right. \nonumber \\ & \left. \tilde{W}_{hh\rightarrow lfh}+2(E_{lfh}+\Delta E_{lfh\rightarrow lfh})\tilde{W}_{lfh\rightarrow lfh} \right] / \nonumber \\ &( \tilde{W}_{ff\rightarrow lfh}+\tilde{W}_{hh\rightarrow lfh}+2\tilde{W}_{lfh\rightarrow lfh})-<E>\label{act_en}
\end{align}
where 
\begin{align}
&<E>=\nonumber \\
&\frac{E_{ff}\exp(-\beta E_{ff})+E_{hh}\exp(-\beta E_{hh})+E_{lfh}\exp(-\beta E_{lfh})}{\exp(-\beta E_{ff})+\exp(-\beta E_{hh})+\exp(-\beta E_{lfh})}
\end{align}
is mean value of dimer energy in the equilibrium conditions. The first term in \ref{act_en} is the weighted arithmetic mean of the energies at the saddle points over which the dimer jumps while changing its orientation. The second term is the average energy that the dimer has at a given temperature. We show all activation energies for Cu(111) surface in the  top panel of Fig. \ref{eff_act_en}. We see that the activation energy for the dimer translational motion is the largest one and this for dimer rotation is lower even than this of the monomer diffusion. Therefore dimer motion consists mostly of rotation, while the probability for the move of the center of mass is much lower. Rotation is realized via intracell jumps and we can compare the calculated  value  $E_a$=20 meV with measured at low temperatures $E_a$  intracell diffusion as 18$\pm$3 meV \cite{Repp1}. Monomer diffusion is somewhere in between. We can also see how prefactors for translational dimer motion (Fig. \ref{eff_pref}) and rotational motion (Fig. \ref{eff_pref_rot}) depend on temperature. It can be seen that prefactor for dimer diffusion is higher than that for monomer, but decreases with temperature in the same way, while prefactor for rotational diffusion increases with temperature reaching constant value from below. Below we will compare this behavior with this on Ag surface.
\begin{figure}
\begin{center}
\includegraphics[angle=-90,width=6.2cm]{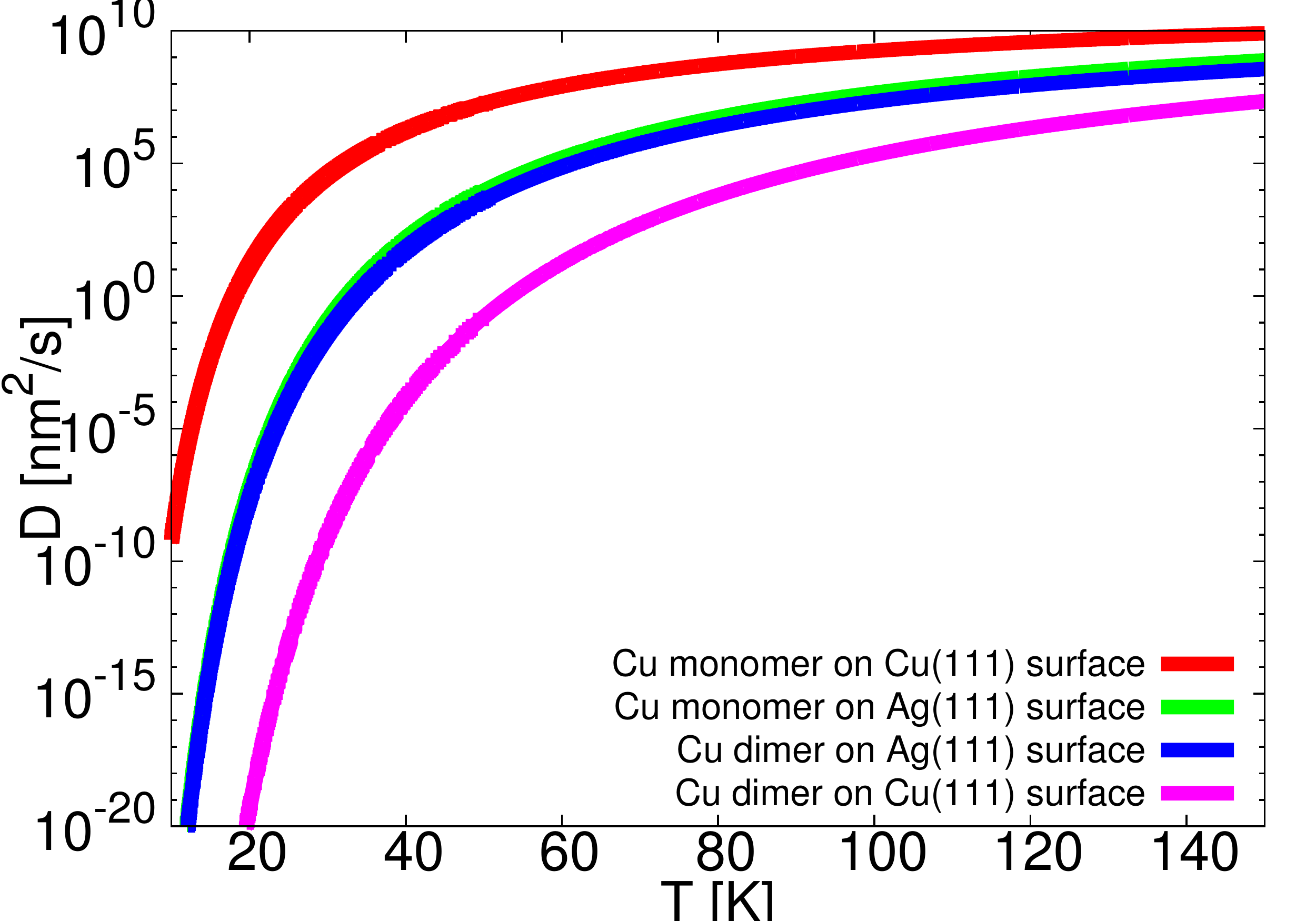}
\includegraphics[angle=-90,width=6.2cm]{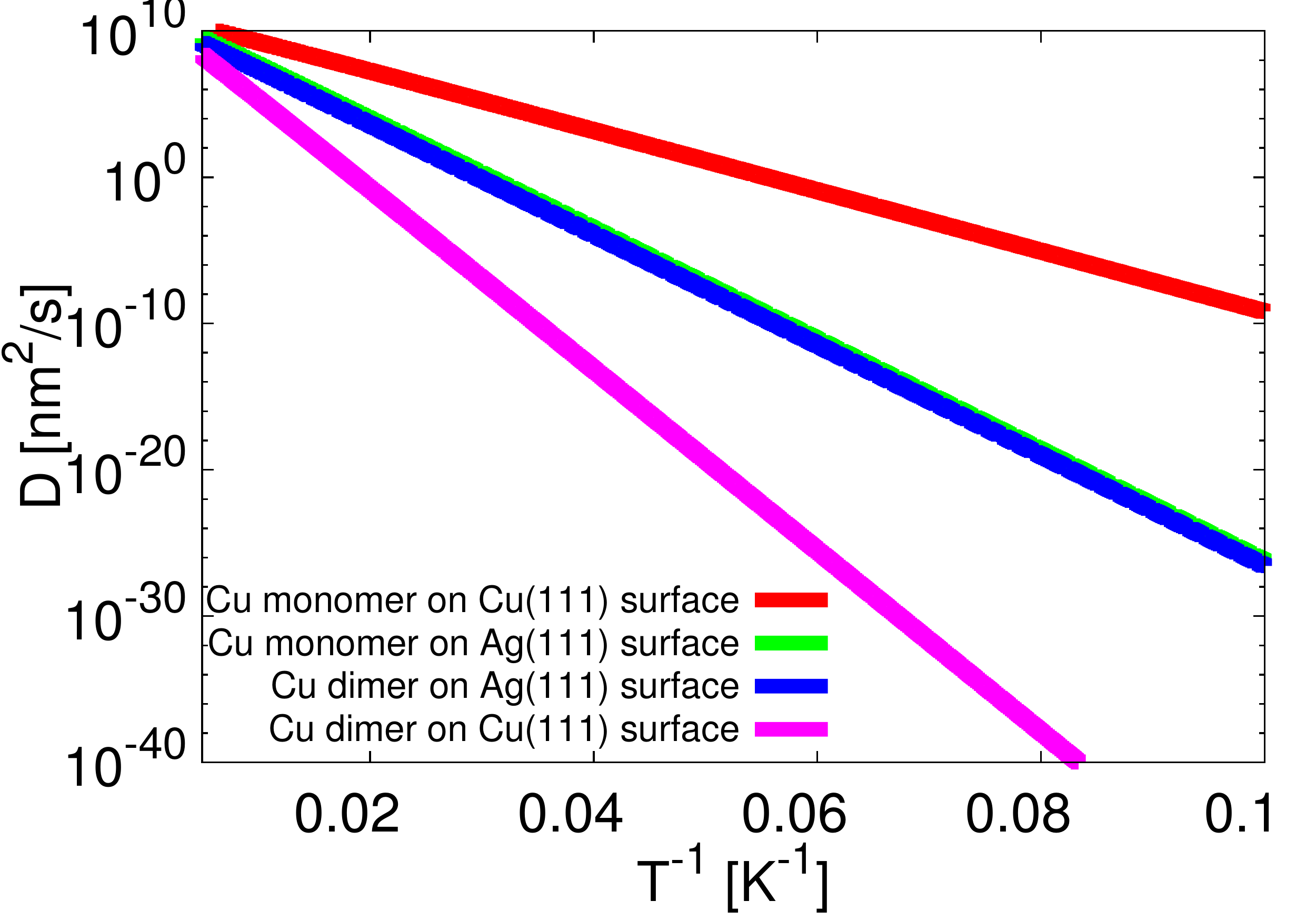}
\end{center}
\caption{Logarithmic scale plot of diffusion coefficient as a function of temperature (left) and  as a function of inverse temperature  (right) for Cu monomers and dimers on the Cu(111) and Ag(111) surfaces at low temperatures. Curves are described from top down. Two lines in the middle show monomer and dimer at Ag surface $\nu=10^{13}/s$.}
\label{diff_coeff_log}
\end{figure}

\subsubsection{Cu dimer on Ag(111) lattice}
For Cu dimer on Ag(111) there is one additional lattice position, namely the $sfh$ site (short $fh$), which wasn't present in the homoepitaxial case. According to Ref. \cite{Hayat1} there are four types of transitions between Cu/Ag(111) sites. The geometry of jumps on this lattice is significantly different than that on Cu/Cu(111) and it requires a separate consideration. The map of the energy minima and the possible jumps in the system is shown in  Fig.~\ref{jumps2}. The path along which we plot energy potential is shown in top panel of Fig.~\ref{jumps2}. It can be seen that it leads from one cell to the next one. As before three neighboring cells can be seen in Fig~\ref{jumps2}. Three $lfh$ sites in each cell have the same position of the center of mass despite being drawn separately and the dashed line between  $ff$ and $hh$ sites represents the direct transition between these sites.

We should point out some differences between the homoepitaxial and the heteroepitaxial case. First of all, as we have already mentioned, $sfh$ position is added. It means that the direct intercell $ff\rightarrow hh$ jump is replaced by jumps through the $sfh$ site (Fig.~\ref{intercell}). Because of this there exist more sequences of jumps that are possible, since the dimer from the $sfh$ site can jump either to one of the two $ff$ sites or to one of the $hh$ sites. Therefore, it is difficult to state unambiguously which jumps are intracell and which are intercell. Another difference is  lack of  $hh\rightarrow lfh$ and $lfh\rightarrow lfh$ connections.

We apply the variational method to the hetoroepitaxial system in the same way as we did for the homoepitaxial one. First we assign geometrical phases to each of the sites taking into account the symmetry, which is the same as in the previous case. Then we insert  phases into  Eq.~(\ref{var_form}) and minimize this expression with respect to them. Resulting formulas for phases can be found in Appendix B. And final expression for diffusion coefficient is
\begin{align}
D&=3a^{2}\{\tilde{W}_{ff\rightarrow hh}\tilde{W}_{ff\rightarrow sfh}\tilde{W}_{hh\rightarrow sfh}(\tilde{W}_{ff\rightarrow sfh}+\tilde{W}_{hh\rightarrow sfh})\nonumber\\
&+\frac{1}{2}\tilde{W}_{ff\rightarrow sfh}\tilde{W}_{hh\rightarrow sfh}(\tilde{W}_{ff\rightarrow hh}\tilde{W}_{ff\rightarrow lfh}\nonumber\\
&+\tilde{W}_{ff\rightarrow lfh}\tilde{W}_{ff\rightarrow sfh}+\tilde{W}_{ff\rightarrow lfh}\tilde{W}_{hh\rightarrow sfh} \nonumber \\&+\tilde{W}_{ff\rightarrow sfh}\tilde{W}_{hh\rightarrow sfh})+\frac{1}{4}\tilde{W}_{ff\rightarrow hh}\tilde{W}_{ff\rightarrow lfh}(\tilde{W}_{ff\rightarrow sfh}^{2}\nonumber\\
&+\tilde{W}_{hh\rightarrow sfh}^{2})\} /\{2\tilde{W}_{ff\rightarrow sfh}\tilde{W}_{hh\rightarrow sfh}(\tilde{W}_{ff\rightarrow hh}+ \tilde{W}_{ff\rightarrow lfh}\nonumber \\ &+\tilde{W}_{ff\rightarrow sfh}+\tilde{W}_{hh\rightarrow sfh})+\tilde{W}_{ff\rightarrow hh}(\tilde{W}_{ff\rightarrow sfh}^{2}\nonumber\\
&+\tilde{W}_{hh\rightarrow sfh}^{2}+\tilde{W}_{ff\rightarrow lfh}\tilde{W}_{ff\rightarrow sfh}+\tilde{W}_{ff\rightarrow lfh}\tilde{W}_{hh\rightarrow sfh}) \nonumber \\ &+\frac{3}{2}\tilde{W}_{ff\rightarrow lfh}\tilde{W}_{hh\rightarrow sfh}^2\}.
\label{15}
\end{align}
The above formula is quite complex, however, there are still a few things that can be deduced from it. First of all, we can see that unlike  the Cu(111) surface, there is no single type of jump on Ag(111) that is crucial for the diffusion. On the other hand, the diffusion needs at least two different types of jumps in order to occur. One single type of jump is unable to trigger diffusion by itself. However, it seems that the $ff\rightarrow sfh$ jump is the most important one since it appears in almost all the diffusion paths. As we can see, the distinction between intracell and intercell jumps is not so obvious here as it was on the Cu(111) surface.

When we ignore the two slowest jumps i.e. $W_{ff\rightarrow hh}=W_{hh\rightarrow sfh}=0$ the  diffusion coefficient formula simplifies to
\begin{equation}
D=\frac{3}{4}a^2P_{ff}^{eq}\frac{W_{ff\rightarrow sfh}W_{ff\rightarrow lfh}}{W_{ff\rightarrow sfh}+W_{ff\rightarrow lfh}}.
\label{16}
\end{equation}
It corresponds to the zigzag motion observed by STM in Ref. \cite{Morgenstern1}.

For a more complex case, where the least probable transition is still $W_{ff\rightarrow hh}=0$ but $W_{hh\rightarrow sfh}\neq 0$,
the diffusion coefficient  is
\begin{align}
&D=3a^2 \left[ \tilde{W}_{ff\rightarrow sfh}\left(\tilde{W}_{ff\rightarrow lfh}
\tilde{W}_{hh\rightarrow sfh}+\tilde{W}_{ff\rightarrow sfh}\tilde{W}_{ff\rightarrow lfh} \right. \right. \nonumber \\ 
& \left.\left. +\tilde{W}_{ff\rightarrow sfh}\tilde{W}_{hh\rightarrow sfh} \right) \right] / \left[ 3\tilde{W}_{ff\rightarrow lfh}\tilde{W}_{hh \rightarrow sfh} \right. \nonumber \\
& \left.+4\tilde{W}_{ff\rightarrow sfh} \left( \tilde{W}_{ff\rightarrow sfh}+\tilde{W}_{ff\rightarrow lfh}+\tilde{W}_{hh\rightarrow sfh}\right) \right]
\label{17}
\end{align}
Here, in addition to the zigzag motion, we have also two other ones. It is also seen that jumps $W_{ff\rightarrow lfh}$ and $W_{hh\rightarrow sfh}$ by themselves are not able to make the dimer diffuse along the surface. It is understandable because these jumps are between two completely different pairs of states. In the bottom panel of Fig. \ref{diff_coeff} we plotted both  approximations Eqs. (\ref{16},\ref{17}). In  Fig. \ref{path_dimer} the successive dimer  paths are shown.  It can be seen that  it is not zigzag motion (plotted in the lowest dashed line) alone that is responsible for the dimer diffusion. With two more modes added to this simplest one we can reproduce diffusion curve more precisely. The most probable dimer movement can be quite complicated as shown in Fig.\ref{path_dimer}.

The effective activation energy in the most general case is expressed by a complicated formula, therefore we do not write the equation, but just plot the results in Fig. \ref{eff_act_en}. Its behavior is a bit more complex than for the monomer's case, what is not surprising since the dimer's map of transitions is also more complex as is the diffusion coefficient. At low temperatures the activation energy for the dimer is 75 meV compared to 73 meV in the experiment \cite{Morgenstern1} and the molecular dynamics calculations \cite{Hayat1}. Between 0 and 100 K activation energy for dimer diffusion is higher than this for monomers and for higher temperatures $E_a$ for dimers decreases below value of  $E_a$ for monomers. Effective prefactor for the dimer plotted in Fig.~\ref{eff_pref} is lower than this for monomers what compensates the difference in activation energy. At lower temperatures where $E_a$ for dimers is higher than that for monomers both diffusion coefficients have the same value as seen in Fig. \ref{diff_coeff_log}. This is different than at Cu surface  where even at low temperatures dimer diffusion coefficient is much lower than this for monomers. It can be also seen that at low temperatures dimers at Ag surface move much faster than at Cu surface. The same diffusion coefficient is shown in the  right panel of Fig.~\ref{diff_coeff_log}. In this scale all lines are straight. Each curve has different slope, given by  activation energy. However, it is impossible to see any difference in activation energies as a function of temperature in such plot.

The rotational diffusion coefficient for the Cu dimer on Ag(111) lattice is equal to
\begin{equation}
D_r=\frac{\pi^2}{18}(\tilde{W}_{ff\rightarrow lfh}+\tilde{W}_{ff\rightarrow sfh}+\tilde{W}_{hh\rightarrow sfh})
\end{equation}
It implies that the rotational diffusive motion in our case is a sum of three parallel motions, each of them  related to a different type of jump. Transition   $W_{ff\rightarrow hh}$ is not present in the expression, which is understandable because this transition does not contribute to the change of the dimer's orientation as do the other three transitions. In contrast to the translational diffusion, the rotational one requires just one type of jump in order to occur. It is with agreement with a simple analysis of the dimer's angular motion on the lattice. Each of the jumps whose rates appear in the expression can by itself rotate the dimer by the full angle. As was in the case of Cu/Cu(111) it is not possible to compare diffusion coefficients for different types of motion (translational and rotational) directly, however, as in previous case in Fig. \ref{diff_rot_coeff} we plotted value $D_r$ multiplied by square of the half of dimer length. It can be seen that rotation on Ag  surface is much slower than rotation on Cu and it is very close to the translational diffusion of dimers on Ag. It can be understood that particles rotate while they move forward.

As was in the case of the translational diffusion, we can also calculate the effective activation energy for the rotational diffusion
\begin{align}
&E_{a}=[(E_{ff}+\Delta E_{ff\rightarrow lfh})\tilde{W}_{ff\rightarrow lfh}+(E_{ff}+\Delta E_{ff\rightarrow sfh})\nonumber \\ &\tilde{W}_{ff\rightarrow sfh} 
+(E_{hh}+\Delta E_{hh\rightarrow sfh})\tilde{W}_{hh\rightarrow sfh}]
\nonumber \\ &/[\tilde{W}_{ff\rightarrow lfh}+\tilde{W}_{ff\rightarrow sfh}+\tilde{W}_{hh\rightarrow sfh}] -< E>
\end{align}
where $<E>$ is mean energy of the dimer in the equilibrium state averaged over all four possible states. Note that even if not all transitions on the lattice are engaged in the rotational diffusion, the energy average is taken over all possible energies in the system.

Because there are three parallel paths of rotational diffusion, the dimer can choose any of them independently. At low temperatures the effective activation energy for the rotational diffusion is 72 meV, which is also the energy barrier for the $ff\rightarrow sfh$ transition. Even though the barrier for the $hh\rightarrow sfh$ transition is even lower, this transition is less favourable because the hh site lies 13 meV higher than the $ff$ site and consequently, the saddle point for the $hh\rightarrow sfh$ jump lies higher than for the $ff\rightarrow sfh$ jump. Therefore, the $ff\rightarrow sfh$ jumps make the easiest diffusion path. The effective activation energy then slightly rises reaching 73 meV at 50 K, possibly due to the other jumps becoming more important. Then above that temperature, as the dimer gets more and more energy, the activation energy starts to fall down very quickly reaching
\begin{align}
E_{a}^{\infty}&=\frac{2E_{ff}+\Delta E_{ff\rightarrow lfh}+\Delta E_{ff\rightarrow sfh}+E_{hh}+\Delta E_{hh\rightarrow sfh}}{3}\nonumber\\
&-\frac{E_{ff}+E_{hh}+E_{sfh}+E_{lfh}}{4}\approx 36 meV
\end{align}
in the  limit of infinite temperature.

In Fig. \ref{eff_pref_rot} we show an effective prefactor for the rotational diffusion. Its behaviour is completely different than that of the translational prefactor. First it rises very quickly and about 50 K reaches a maximum slightly above $6^.10^{12}/s$, which is almost twice as high as the value at 0 K. After that maximum the prefactor goes down, just as in the translational cases.

The different behaviour of both the activation energy and the prefactor for the rotational motion compared to the translational one can be explained due to the fact that the former is more parallel while the latter (especially for the monomer) is more serial.

\section{Conclusions}
Systems of diffusing Cu adatoms and dimers  were compared at (111)  surfaces for homo and heteroepitaxial systems. Cu and Ag crystals have different lattice constants and as a result potential energy surface seen by adatoms varies. However, for both surfaces diffusion of single adatoms is simple and has the same character and values given by appropriate activation energies for these processes. For dimers the situation is completely different. At Ag surface even if the lattice of dimer jumps is  quite complicated, involving all types of possible moves, all jumps are of the same  rate and as  an effect at low temperatures diffusion is quite fast. At higher temperatures it slows down. For homogeneous diffusion of Cu dimers the energy surface is more complex. Jumps inside hexagonal cells are much faster than these between cells. Diffusion either happens only via intercell jumps - this is slower path, or it partly goes via intracell jumps, which is slightly faster. Both diffusion channels together give diffusion coefficient that is lower than this for Ag surface, but at higher temperatures, around 600 K the relation turns back. It is interesting to compare  temperature dependence of rotational movement in both cases. They both have   rather simple form being a sum of independent diffusional modes,  but their effective activation energies are very different. Whereas activation energy for rotational motion at Cu surface is very low and almost independent of temperature, $E_a$ at Ag is higher and decreases with temperature. As an effect rotation at Cu surface is as fast as single particle movement at this surface and it is localized inside cells, whereas rotation at Ag lattice is as fast as dimer movement and it happens simultaneously with translational movement of dimers. At low temperatures it is the same as monomer diffusion at Ag surface. It can be seen that even if both surfaces have similar geometry and not very distinct lattice constants, the character of resulting  monomer and  dimer diffusion is different.

The general consequence of diffusion over complex  energy surface is temperature dependence of activation energy and  prefactor of diffusion coefficient. For such an effect to appear there is no need for   special interactions or other correlations in the system. Interestingly prefactors in our case drastically change on coming from low to high temperatures as shown in Fig 5. Such temperature dependence of prefactors for monomer or dimer diffusion could explain some discrepancies in results obtained   from different experiments \cite{chvoj,chvoj2,bietti}.   Especially change of prefactors ratio for diffusion at step and on terraces  for Ag/Ag(111) system \cite{chvoj,chvoj2} can be considered as an effect of multistate particle diffusion. We have shown above that  depending on the temperature the character of  diffusion on  smooth (111) metallic surfaces changes due to small   differences of the  energy barriers for adatom or dimer jumps. We have also shown that at these surfaces dimer diffusion is more complicated, so it can be regarded as a good candidate for the interpretation of the experimental data.
\section{Acknowledgment}
Research supported  by the National Science Centre(NCN) of Poland
(Grant NCN No. 2013/11/D/ST3/02700)

\section{Appendix A}
Because we deal with an infinite periodic lattice it is convenient to take the Fourier transform of the equation (\ref{master_eq})
\begin{equation}
\frac{d}{dt}P_{\alpha}(\vec{k};t)=\sum_{\gamma\neq\alpha}M(\gamma;\alpha)P_{\gamma}(\vec{k};t)+M(\alpha;\alpha)P_{\alpha}(\vec{k};t),\label{eq_matrix}
\end{equation}
where $P_{\alpha}(\vec{k};t)=\sum_{j}\exp(i\vec{k}\vec{r}_{j}^{\alpha})P(j,\alpha;t)$. Elements $M(\gamma;\alpha)$ are expressed in terms of the jump rates in the following way:
\begin{eqnarray}
M(\alpha;\alpha)=-\sum_{\gamma\neq\alpha}W_{\alpha,\gamma},\nonumber\\
M(\gamma;\alpha)=W_{\gamma,\alpha}e^{i\vec{k}(\vec{r}_{l\rq{}}^{\gamma}-\vec{r}_{l}^{\alpha})}.
\end{eqnarray}
Sum in the first expression is over all possible transitions from the state $\alpha$ to the neighboring states. The second expression is written  for each transition from $\gamma$ to $\alpha$ and $l\rq{}=l$ as long as jump is within chosen cell, whereas $l\rq{}=l\pm1$ when jump of  adparticle transforms it to the neighboring cell. 

The matrix $\hat{M}$ contains all the information about system dynamics. Its eigenvalues correspond to dynamic modes that describe relaxation of the system towards the equilibrium. It can be shown that of all the eigenvalues the diffusive one has the smallest absolute value \cite{haus,Titulaer1,Gortel1}. Therefore, we can use the variational method in order to obtain the diffusion coefficient
\begin{equation}
\lim_{|\vec{k}|\rightarrow 0}\lambda_{D}=\lim_{|\vec{k}|\rightarrow 0}\frac{\vec{w}\hat{M}\vec{v}}{\vec{w}\vec{v}}=-\vec{k}\hat{D}\vec{k},
\end{equation}
where $\vec{w}$ and $\vec{v}$ are left and right trial eigenvectors, respectively. Their elements are related to each other by $w_{\alpha}^{*}P_{eq}(\alpha)=v_{\alpha}$. Assuming $w_{\alpha}=e^{-i\vec{k}\vec{\phi}_{\alpha}}$, where $\vec{\phi}_{\alpha}$ is the geometric phase of the adsorption site $\alpha$, we get the variational formula for the diffusion coefficient in the form
\begin{align}
\vec{k}\hat{D}_{var}\vec{k}&=\lim_{|\vec{k}|\rightarrow 0}\sum_{\alpha>\gamma}W_{\alpha;\gamma}P_{eq}(\alpha)|e^{i\vec{k}(\vec{r}_{l}^{\gamma}+\vec{\phi}_{\gamma})}-e^{i\vec{k}(\vec{r}_{j}^{\alpha}+\vec{\phi}_{\alpha})}|^2\nonumber\\
&=\sum_{\alpha>\gamma}W_{\alpha;\gamma}P_{eq}(\alpha)[\vec{k}(\vec{r}_{l}^{\gamma}+\vec{\phi}_{\gamma}-\vec{r}_{j}^{\alpha}-\vec{\phi}_{\alpha})]^2\label{var_form}.
\end{align}
The above choice of variational vector has been so far valid in the cases of single adatom diffusion. Each of the vector phases corresponds to exactly one adsorption site within a single cell and each of its vector components is coupled either to x or y direction on the surface. The same approach can be used to describe the diffusion of the dimer center of mass. Moreover, the procedure can be also applied for rotational degrees of freedom instead of the translational one in order to calculate rotational diffusion coefficient. 

The total number of variational parameters, whose values should be chosen in such a way that the above expression is minimized, is twice the number of different types of adsorption sites. However, since only the phase differences contribute to the expression, we can always shift all the phases simultaneously by the same value. In such a way we can set one of the vector phases to $\vec{0}$. Moreover, the remaining phases can be related to each other due to the system's symmetry, which leads to further reduction of independent parameters. Ultimately we have a set of several linear equation, one for each independent parameter. We insert solution of these equations to Eq. (\ref{var_form}) and receive final expression for the diffusion coefficient.

\section{Appendix B}
First step in the calculations of diffusion coefficient described above is to find variational parameters from the equation (\ref{var_form}). Phases are associated with sites of the lattice that represents possible jumps in the phase space. Dimer configurations are   defined by  positions of both particles. In the case of  Cu dimer diffusion at Cu(111) surface sites are marked by $ff, lfh$ and  $hh$. Each  site has attributed phase vector of orientation consistent with the lattice symmetry. Final expression for diffusion coefficient given by (\ref{11}) is obtained for  $\delta_{lfh}$ phase which turns out to be  zero and the other two
  \begin{align}
 & \delta_{ff}=-\{\tilde{W}_{ff\rightarrow lfh}^{2}(\frac{\tilde{W}_{ff\rightarrow hh}}{2}+\tilde{V}_{ff\rightarrow hh}) \nonumber \\ &+\tilde{W}_{hh\rightarrow lfh}^{2}(\frac{\tilde{W}_{ff\rightarrow hh}}{2}-2\tilde{V}_{ff\rightarrow hh})\nonumber\\
  &+\tilde{W}_{ff\rightarrow lfh}\tilde{W}_{hh\rightarrow lfh}(\tilde{W}_{ff\rightarrow hh}-\tilde{V}_{ff\rightarrow hh} \nonumber \\
& +\tilde{W}_{ff\rightarrow lfh}+\tilde{W}_{hh\rightarrow lfh}+2\tilde{W}_{lfh\rightarrow lfh})\nonumber\\
  &+\tilde{W}_{ff\rightarrow lfh}\tilde{W}_{lfh\rightarrow lfh}(\tilde{W}_{ff\rightarrow hh}+2\tilde{V}_{ff\rightarrow hh}) \nonumber \\
&+\tilde{W}_{hh\rightarrow lfh}\tilde{W}_{lfh\rightarrow lfh}(\tilde{W}_{ff\rightarrow hh}-4\tilde{V}_{ff\rightarrow hh})\}\nonumber\\
  &/\{[(\tilde{W}_{ff\rightarrow hh}+2\tilde{V}_{ff\rightarrow hh})(\tilde{W}_{ff\rightarrow lfh}+\tilde{W}_{hh\rightarrow lfh}) \nonumber \\
&+2\tilde{W}_{ff\rightarrow lfh}\tilde{W}_{hh\rightarrow lfh}][\tilde{W}_{ff\rightarrow lfh}+\tilde{W}_{hh\rightarrow lfh}\nonumber\\
  &+2\tilde{W}_{lfh\rightarrow lfh}]\}
\end{align}
\begin{align}
  &\delta_{hh}=\{\tilde{W}_{ff\rightarrow lfh}^{2}(\frac{\tilde{W}_{ff\rightarrow hh}}{2}-2\tilde{V}_{ff\rightarrow hh})+\tilde{W}_{hh\rightarrow lfh}^{2}(\frac{\tilde{W}_{ff\rightarrow hh}}{2}\nonumber\\
  &+\tilde{V}_{ff\rightarrow hh})+\tilde{W}_{ff\rightarrow lfh}\tilde{W}_{hh\rightarrow lfh}(\tilde{W}_{ff\rightarrow hh}-\tilde{V}_{ff\rightarrow hh} \nonumber \\ &+\tilde{W}_{ff\rightarrow lfh}+\tilde{W}_{hh\rightarrow lfh}+2\tilde{W}_{lfh\rightarrow lfh})\nonumber\\
  &+\tilde{W}_{ff\rightarrow lfh}\tilde{W}_{lfh\rightarrow lfh}(\tilde{W}_{ff\rightarrow hh}-4\tilde{V}_{ff\rightarrow hh})\nonumber \\ &+\tilde{W}_{hh\rightarrow lfh}\tilde{W}_{lfh\rightarrow lfh}(\tilde{W}_{ff\rightarrow hh}+2\tilde{V}_{ff\rightarrow hh})\}\nonumber\\
  &/\{[(\tilde{W}_{ff\rightarrow hh}+2\tilde{V}_{ff\rightarrow hh})(\tilde{W}_{ff\rightarrow lfh}+\tilde{W}_{hh\rightarrow lfh}) \nonumber \\ &+2\tilde{W}_{ff\rightarrow lfh}\tilde{W}_{hh\rightarrow lfh}][\tilde{W}_{ff\rightarrow lfh}+\tilde{W}_{hh\rightarrow lfh}+2\tilde{W}_{lfh\rightarrow lfh}]\}
  \end{align}

When Cu dimer diffuses on Ag(111) surface it has four different positions $ff, sfh, lfh$ or $hh$.
In the most general case, with all the jumps taken into account  diffusion is given by  Eq. (\ref{15}) where the following  variational parameters were used 
\begin{align}
&\delta_{ff}=-\{\frac{3}{4}\tilde{W}_{ff\rightarrow lfh}\tilde{W}_{hh\rightarrow sfh}^{2}+\frac{3}{2}\tilde{W}_{ff\rightarrow hh}\tilde{W}_{hh\rightarrow sfh}^{2} \nonumber \\
&-\tilde{W}_{ff\rightarrow sfh}^{2}\tilde{W}_{hh\rightarrow sfh}+\tilde{W}_{ff\rightarrow sfh}\tilde{W}_{ff\rightarrow lfh}\tilde{W}_{hh\rightarrow sfh}\nonumber\\
&-\frac{1}{2}\tilde{W}_{ff\rightarrow sfh}^{2}\tilde{W}_{ff\rightarrow hh}+\tilde{W}_{ff\rightarrow sfh}\tilde{W}_{ff\rightarrow hh}\tilde{W}_{hh\rightarrow sfh}\nonumber\\
&+\frac{1}{2}\tilde{W}_{ff\rightarrow lfh}(\tilde{W}_{ff\rightarrow hh}\tilde{W}_{hh\rightarrow sfh}+\tilde{W}_{ff\rightarrow sfh}\tilde{W}_{ff\rightarrow hh})\}\nonumber\\
&/\{2\tilde{W}_{ff\rightarrow sfh}\tilde{W}_{hh\rightarrow sfh}^{2}+\frac{3}{2}\tilde{W}_{ff\rightarrow lfh}\tilde{W}_{hh\rightarrow sfh}^{2}  \nonumber \\
&+\tilde{W}_{ff\rightarrow hh}\tilde{W}_{hh\rightarrow sfh}^{2}+2\tilde{W}_{ff\rightarrow sfh}^{2}\tilde{W}_{hh\rightarrow sfh}\nonumber\\
&+\tilde{W}_{ff\rightarrow sfh}^{2}\tilde{W}_{ff\rightarrow hh}+2\tilde{W}_{ff\rightarrow sfh}\tilde{W}_{ff\rightarrow lfh}\tilde{W}_{hh\rightarrow sfh} \nonumber \\ &
+2\tilde{W}_{ff\rightarrow sfh}\tilde{W}_{ff\rightarrow hh}\tilde{W}_{hh\rightarrow sfh}+\tilde{W}_{ff\rightarrow lfh}\tilde{W}_{ff\rightarrow hh}\tilde{W}_{hh\rightarrow sfh}\nonumber\\
&+\tilde{W}_{ff\rightarrow sfh}\tilde{W}_{ff\rightarrow lfh}\tilde{W}_{ff\rightarrow hh}\}
\end{align}
\begin{align}
&\delta_{hh}=-\{\frac{3}{4}\tilde{W}_{ff\rightarrow lfh}\tilde{W}_{hh\rightarrow sfh}^{2}+\frac{1}{2}\tilde{W}_{ff\rightarrow hh}\tilde{W}_{hh\rightarrow sfh}^{2}\nonumber\\
&
+\tilde{W}_{ff\rightarrow sfh}\tilde{W}_{hh\rightarrow sfh}^{2}+\frac{1}{2}\tilde{W}_{ff\rightarrow sfh}\tilde{W}_{ff\rightarrow lfh}\tilde{W}_{hh\rightarrow sfh} \nonumber \\ &-\frac{3}{2}\tilde{W}_{ff\rightarrow sfh}^{2}\tilde{W}_{ff\rightarrow hh}
-\tilde{W}_{ff\rightarrow sfh}\tilde{W}_{ff\rightarrow hh}\tilde{W}_{hh\rightarrow sfh}\nonumber\\
&-\frac{1}{2}\tilde{W}_{ff\rightarrow lfh}(\tilde{W}_{ff\rightarrow hh}\tilde{W}_{hh\rightarrow sfh}-\tilde{W}_{ff\rightarrow sfh}\tilde{W}_{ff\rightarrow hh})\}\nonumber\\
&/\{2\tilde{W}_{ff\rightarrow sfh}\tilde{W}_{hh\rightarrow sfh}^{2}+\frac{3}{2}\tilde{W}_{ff\rightarrow lfh}\tilde{W}_{hh\rightarrow sfh}^{2}  \nonumber \\
&+\tilde{W}_{ff\rightarrow hh}\tilde{W}_{hh\rightarrow sfh}^{2}+2\tilde{W}_{ff\rightarrow sfh}^{2}\tilde{W}_{hh\rightarrow sfh}\nonumber\\
&+\tilde{W}_{ff\rightarrow sfh}^{2}\tilde{W}_{ff\rightarrow hh}+2\tilde{W}_{ff\rightarrow sfh}\tilde{W}_{ff\rightarrow lfh}\tilde{W}_{hh\rightarrow sfh} \nonumber \\ &
+2\tilde{W}_{ff\rightarrow sfh}\tilde{W}_{ff\rightarrow hh}\tilde{W}_{hh\rightarrow sfh}+\tilde{W}_{ff\rightarrow lfh}\tilde{W}_{ff\rightarrow hh}\tilde{W}_{hh\rightarrow sfh}\nonumber\\
&+\tilde{W}_{ff\rightarrow sfh}\tilde{W}_{ff\rightarrow lfh}\tilde{W}_{ff\rightarrow hh}\}
\end{align}
\begin{align}
\delta_{sfh}=&\frac{\tilde{W}_{ff\rightarrow hh}}{\tilde{W}_{hh\rightarrow sfh}}\left(1+\delta_{ff}-\delta_{hh}\right)-2\delta_{hh}-\frac{1}{2}\nonumber\\
\delta_{lfh}=&-\frac{1}{4}-\frac{\delta_{ff}}{2}
\end{align}

For the case $W_{ff\rightarrow hh}=W_{hh\rightarrow sfh}=0$ the phases simplify to
\begin{align}
\delta_{ff}&=\frac{\tilde{W}_{ff\rightarrow sfh}-\tilde{W}_{ff\rightarrow lfh}}{2(\tilde{W}_{ff\rightarrow sfh}+\tilde{W}_{ff\rightarrow lfh})}\nonumber\\
\delta_{hh}&=0\nonumber\\
\delta_{sfh}&=\delta_{lfh}=-\frac{\tilde{W}_{ff\rightarrow sfh}}{2(\tilde{W}_{ff\rightarrow sfh}+\tilde{W}_{ff\rightarrow lfh})}
\end{align}
and we obtain Eq. (\ref{16}).
For a more complex case, where the least probable transition is still $W_{ff\rightarrow hh}=0$ but $W_{hh\rightarrow sfh}\neq 0$, we have the following values of phases
\begin{align}
&\delta_{ff}=\left[-\frac{3}{4}\tilde{W}_{ff\rightarrow lfh}\tilde{W}_{hh\rightarrow sfh}+\tilde{W}_{ff\rightarrow sfh}^{2}\right.\nonumber\\&\left.-\tilde{W}_{ff\rightarrow sfh}\tilde{W}_{ff\rightarrow lfh}\right]
/ \left[\frac{3}{4}\tilde{W}_{ff\rightarrow lfh}\tilde{W}_{hh\rightarrow sfh}\right.\nonumber \\
&\left.+2\tilde{W}_{ff\rightarrow sfh}(\tilde{W}_{hh\rightarrow sfh}+\tilde{W}_{ff\rightarrow sfh}+\tilde{W}_{ff\rightarrow lfh}) \right]
\end{align}
\begin{align}
&\delta_{hh}=-\left[\frac{3}{4}\tilde{W}_{ff\rightarrow lfh}\tilde{W}_{hh\rightarrow sfh}+\tilde{W}_{ff\rightarrow sfh}\tilde{W}_{hh\rightarrow sfh}\right.\nonumber \\
&\left.+\frac{1}{2}\tilde{W}_{ff\rightarrow sfh}\tilde{W}_{ff\rightarrow lfh}\right] / \left[\frac{3}{2}\tilde{W}_{ff\rightarrow lfh}\tilde{W}_{hh\rightarrow sfh}\right.\nonumber\\&\left.+2\tilde{W}_{ff\rightarrow sfh}(\tilde{W}_{hh\rightarrow sfh}+\tilde{W}_{ff\rightarrow sfh}+\tilde{W}_{ff\rightarrow lfh}) \right]
\end{align}
\begin{align}
\delta_{sfh}&=-\frac{1}{2}-2\delta_{hh}\nonumber\\
\delta_{lfh}&=-\frac{1}{4}-\frac{\delta_{ff}}{2}
\end{align}
and these phases  lead to Eq.(\ref{17})

\end{document}